\documentclass[prd,showkeys,floatfix,twocolumn,amsmath,amssymb,floatfix]{revtex4}
\usepackage{amsmath}
\usepackage{graphicx}
\usepackage{float}
\usepackage{epstopdf}
\usepackage{multirow}
\usepackage{subfigure}
\usepackage{color}
\usepackage[colorlinks,citecolor=blue,anchorcolor=red,menucolor=red,linkcolor=red,filecolor=red,runcolor=red,urlcolor=blue,frenchlinks=red]{hyperref}

\allowdisplaybreaks[3]

\begin{document}
\title{Strong decay properties of P-wave single bottom baryons
\\
of the SU(3) flavor antitriplet $\bf\bar 3_F$ }
%
\author{Yi-Jie Wang$^1$}
\author{Xuan Luo$^1$}
\email{cnxluo@seu.edu.cn}
\author{Hua-Xing Chen$^1$}
\email{hxchen@seu.edu.cn}
\author{Er-Liang Cui$^2$}
\email{erliang.cui@nwafu.edu.cn}
\author{Wei-Han Tan$^1$}
\email{tanwh@seu.edu.cn}
\author{Zhi-Yong Zhou$^1$}
\email{zhouzhy@seu.edu.cn}
\affiliation{
$^1$School of Physics, Southeast University, Nanjing 210094, China
\\
$^2$College of Science, Northwest A\&F University, Yangling 712100, China
}

\begin{abstract}
We study the $P$-wave bottom baryons of the $SU(3)$ flavor antitriplet and systematically calculate their strong decay properties, including their $D$-wave decays into ground-state bottom baryons with light pseudoscalar mesons and $S$-wave decays into ground-state bottom baryons with light vector mesons. Together with Refs.~\cite{Tan:2023opd,Yang:2019cvw,Yang:2020zrh,Luo:2024jov}, a rather complete investigation has been performed to study their mass spectra and strong/radiative decay properties, through the methods of QCD sum rules and light-cone sum rules within the framework of heavy quark effective theory. Among various possibilities, we identify four $\Lambda_b$ and four $\Xi_b$ baryons, with limited decay widths and so capable of being observed in experiments. Their masses, mass splittings within the same multiplets, and strong/radiative decay widths are summarized in Table~\ref{tab:decayb3f} for future experimental searching.
\end{abstract}
\maketitle
\pagenumbering{arabic}
%
%
%
\section{Introduction}\label{sec:intro}
%
The single bottom baryon is a three-quark system consisting of one heavy $bottom$ quark and two light $up/down/strange$ quarks~\cite{Gell-Mann:1964ewy,Zweig:1964ruk,Ebert:1996ec,Gerasyuta:1999pc,Karliner:2008sv}. The investigation of this system can provide crucial insights into the understanding of the non-perturbative QCD in the low energy region. Nowadays, all the ground-state bottom baryons, except the $\Omega_b^*$ baryon of $J^P=3/2^+$, have been observed and well identified in experiments~\cite{PDG:2020ssz}. Besides, a significant number of excited bottom baryons have been reported, for example:
\begin{itemize}

\item
In 2012 the excited bottom baryons $\Lambda_b(5912)^0$ and $\Lambda_b(5920)^0$ were observed by LHCb~\cite{LHCb:2012kxf} and CDF~\cite{CDF:2013pvu} in the $\Lambda_b^0\pi^+\pi^-$ invariant mass spectrum. Their masses and widths were measured to be~\cite{PDG2024}:
\begin{eqnarray}
\Lambda_b(5912)^0 &:& M = 5912.19 \pm 0.17{\rm~MeV} \, ,
 \nonumber
\\      && \Gamma < 0.25{\rm~MeV}~{\rm at}~90\%~{\rm CL} \, ;
\\ \Lambda_b(5920)^0 &:& M = 5920.09 \pm 0.17{\rm~MeV} \, ,
\nonumber
\\      && \Gamma < 0.19{\rm~MeV}~{\rm at}~90\%~{\rm CL} \, .
\end{eqnarray}

\item
In 2021 the CMS collaboration observed the $\Xi_b(6100)^-$ in the $\Xi_b^-\pi^+\pi^-$ invariant mass spectrum~\cite{CMS:2021rvl}. Later in 2023 the LHCb collaboration observed the $\Xi_b(6087)^0$ and $\Xi_b(6095)^0$ in the $\Xi_b^0 \pi^+\pi^-$ invariant mass spectrum~\cite{LHCb:2023zpu}. The $\Xi_b(6100)^-$ and $\Xi_b(6095)^0$ are probably isospin partners. Their masses and widths were measured to be~\cite{PDG2024}:
\begin{eqnarray}
\nonumber
\Xi_b(6100)^- &:& M = 6100.3 \pm 0.2 \pm 0.1 \pm 0.6{\rm~MeV} \, ,
\\       && \Gamma =0.94\pm0.30\pm0.08{\rm~MeV} \, ;
\\\nonumber
\Xi_b(6095)^0 &:& M = 6095.4 \pm 0.15 \pm 0.03 \pm 0.5{\rm~MeV} \, ,
\\      && \Gamma =0.50\pm0.33\pm0.11{\rm~MeV} \, ;
\\\nonumber
\Xi_b(6087)^0 &:& M = 6087.24 \pm 0.20 \pm 0.06 \pm 0.5{\rm~MeV} \, ,
\\      && \Gamma =2.43\pm0.51\pm0.10{\rm~MeV} \, .
\end{eqnarray}

\end{itemize}
These states are good candidates for the $P$-wave bottom baryons of the $SU(3)$ flavor antitriplet. Their detailed discussions can be found in the reviews~\cite{Liu:2024uxn,Crede:2013kia,Chen:2016spr,Cheng:2021qpd,Chen:2022asf}. Based on these observations, extensive theoretical investigations have been performed through a variety of phenomenological models, including various quark models~\cite{Garcilazo:2007eh,Ebert:2007nw,Roberts:2007ni,Ortega:2012cx,Yoshida:2015tia,Nagahiro:2016nsx,Wang:2018fjm,Gutierrez-Guerrero:2019uwa,Kawakami:2019hpp,Xiao:2020oif,He:2021xrh,Wang:2017kfr}, various hadronic molecular models~\cite{GarciaRecio:2012db,Liang:2014eba,An:2017lwg,Montana:2017kjw,Debastiani:2017ewu,Chen:2017xat,Nieves:2017jjx,Huang:2018bed,Nieves:2019jhp,Liang:2020dxr,Huang:2018wgr,Yu:2018yxl}, the quark pair creation model~\cite{Chen:2018vuc,Yang:2018lzg,Liang:2020hbo}, QCD sum rules~\cite{Mao:2015gya,Aliev:2018vye,Aliev:2018lcs,Wang:2020pri,Chen:2015kpa,Tan:2023opd,Chen:2017sci,Yang:2019cvw,Yang:2020zrh,Yang:2022oog,Liu:2007fg,Cui:2019dzj,Chen:2020mpy,Luo:2024jov}, Lattice QCD~\cite{Padmanath:2013bla,Burch:2015pka,Padmanath:2017lng,Can:2019wts}, and the chiral perturbation theory~\cite{Cheng:2006dk,Lu:2014ina,Cheng:2015naa}, etc.

\begin{figure}[hbtp]
\begin{center}
\includegraphics[width=0.25\textwidth]{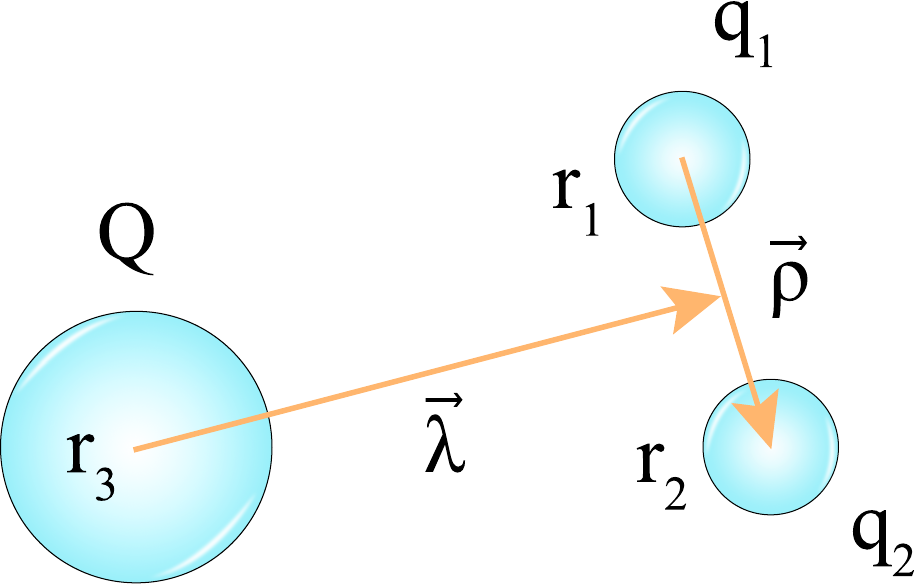}
\end{center}
\caption{Jacobi coordinates $\vec \lambda$ and $\vec \rho$ for the singly bottom baryon system.}
\label{fig:Jacobi}
\end{figure}

We have studied the $P$-wave bottom baryons within the framework of heavy quark effective theory~\cite{Eichten:1989zv,Grinstein:1990mj,Falk:1990yz} in a series of works:
\begin{itemize}

\item In Ref.~\cite{Mao:2015gya} we applied the QCD sum rule method~\cite{Gimenez:2005nt,Shifman:1978bx,Shifman:1978by,Reinders:1984sr,Narison:2002woh,Nielsen:2009uh,Gubler:2018ctz} to systematically calculate their mass spectrum.

\item In Ref.~\cite{Tan:2023opd} we applied the light-cone sum rule method~\cite{Braun:1988qv,Chernyak:1990ag,Ball:1998je,Ball:2006wn,Ball:2004rg,Ball:1998kk,Ball:1998sk,Ball:1998ff,Ball:2007rt,Ball:2007zt} to partly calculate their strong decay properties, including their $S$-wave decays into ground-state bottom baryons with light pseudoscalar mesons.

\item In Ref.~\cite{Luo:2024jov} we applied the light-cone sum rule method to systematically calculate their radiative decay properties.

\end{itemize}
Our results suggest that the $\Lambda_b(5912)^0$, $\Lambda_b(5920)^0$, $\Xi_b(6087)^0$, and $\Xi_b(6095)^0/\Xi_b(6100)^-$ can be well explained as the $P$-wave bottom baryons of the $\rho$-mode, where the orbital excitation is between the two light quarks, as depicted in Fig.~\ref{fig:Jacobi}. This interpretation differs from the $\lambda$-mode interpretation supported by the quark model calculations~\cite{Yoshida:2015tia,Nagahiro:2016nsx,Wang:2017kfr,Kawakami:2019hpp,He:2021xrh}, where the orbital angular momentum is between the bottom quark and the light diquark. Hence, further experimental and theoretical studies are necessary to improve our understanding on their internal structure.

In this paper we shall investigate the $P$-wave bottom baryons of the $SU(3)$ flavor antitriplet  and systematically calculate their strong decay properties, including their $D$-wave decays into ground-state bottom baryons with light pseudoscalar mesons and $S$-wave decays into ground-state bottom baryons with light vector mesons. Together with Refs.~\cite{Mao:2015gya,Tan:2023opd,Luo:2024jov}, we shall arrive at a rather complete investigation on the $P$-wave bottom baryons of the $SU(3)$ flavor antitriplet, allowing us to understand them as a whole. Note that similar investigations have been performed in Refs.~\cite{Yang:2019cvw,Yang:2020zrh} to study the $P$-wave charmed and bottom baryons of the $SU(3)$ flavor sextet.

This paper is organized as follows. In Sec.~\ref{sec:sumrule} we briefly introduce our notations to describe the $P$-wave bottom baryons of the $SU(3)$ flavor antitriplet. In Sec.~\ref{sec:decay} we study their strong decay properties through the light-cone sum rule method. In Sec.~\ref{sec:summary} we discuss the obtained results and conclude this paper.

%
\section{Mass spectrum}
\label{sec:sumrule}

\begin{figure*}[hbtp]
\begin{center}
\includegraphics[width=0.8\textwidth]{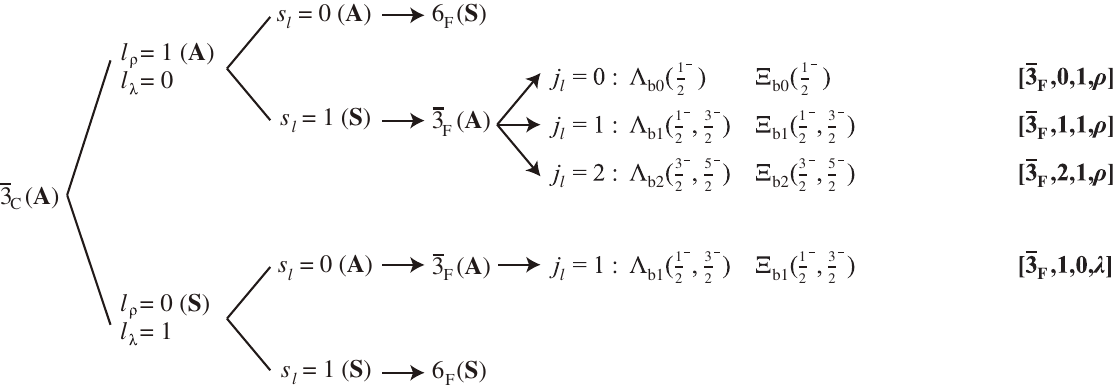}
\end{center}
\caption{Categorization of the $P$-wave bottom baryons belonging to the $SU(3)$ flavor $\mathbf{\bar 3}_F$ representation.}
\label{fig:pwave}
\end{figure*}

In this section we briefly introduce our notations. A single bottom baryon is composed of one heavy $bottom$ quark and two light $up/down/strange$ quarks. To well understand this system, we need to meticulously analyze the symmetries between the two light quarks, including the color, flavor, spin, and orbital degrees of freedom:
\begin{itemize}

\item The two light quarks always have the antisymmetric color structure $\mathbf{\bar 3}_C$.

\item Their $SU(3)$ flavor structure is either symmetric ($\mathbf{6}_F$) or antisymmetric ($\mathbf{\bar 3}_F$).

\item Their spin structure is either symmetric ($s_l = 1$) or antisymmetric ($s_l = 0$), with $s_l$ the total spin of the two light quarks.

\item Their orbital structure is either symmetric ($l_\rho = 0/2/4/\cdots$) or antisymmetric ($l_\rho = 1/3/5/\cdots$), with $l_\rho$ the orbital angular momentum between the two light quarks, as depicted in Fig.~\ref{fig:Jacobi}. Besides, we use $l_\lambda$ to denote the orbital angular momentum between the bottom quark and the light diquark. There may exist both the $\rho$-mode $P$-wave bottom baryons ($l_\rho = 1$ and $l_\lambda=0$) and $\lambda$-mode ones ($l_\rho = 0$ and $l_\lambda=1$).

\end{itemize}
In the context of the Pauli principle, the total symmetry of the two light quarks is antisymmetric. Consequently, we can categorize the $P$-wave bottom baryons into eight multiplets, four of which belong to the $SU(3)$ flavor antitriplet representation, as illustrated in Fig.~\ref{fig:pwave}. We denote them as $[\text{multiplet}, j_l, s_l, \rho/\lambda]$, with $j_l = l_\rho \otimes l_\lambda \otimes s_l$ the total angular momentum of the light component. Each multiplet contains one or two bottom baryons, having the total angular momenta $J = {j_l} \otimes {s_Q} = \left| {{j_l} \pm 1/2} \right|$.

The mass spectrum of the $P$-wave bottom baryons belonging to the $SU(3)$ flavor antitriplet  representation has been systematically studied in Ref.~\cite{Tan:2023opd} through the QCD sum rule method within the framework of heavy quark effective theory, as summarized in Table~\ref{tabmass}. In the next section we shall use these parameters to study their strong decay properties through the light-cone sum rule method still within the framework of heavy quark effective theory. Besides, we need the following parameters for the ground-state bottom baryons and light pseudoscalar/vector mesons, which are taken from PDG~\cite{PDG2024}:
\begin{itemize}

\item For the ground-state bottom baryons, we use
\begin{eqnarray}
\nonumber\Lambda_b(1/2^+)&:& M=5619.60~{\rm MeV}\, ,
\\ \nonumber \Xi_b(1/2^+)&:& M=5793.20~{\rm MeV}\, ,
\\ \nonumber \Sigma_b(1/2^+)&:& M=5813.4~{\rm MeV}\, ,
\\ \nonumber \Sigma^*_b(3/2^+)&:& M=5833.6~{\rm MeV}\, ,
\\ \nonumber \Xi^\prime_b(1/2^+)&:& M=5935.02~{\rm MeV}\, ,
\\ \nonumber \Xi^*_b(3/2^+)&:& M=5952.6~{\rm MeV}\, ,
\\ \nonumber \Omega_b(1/2^+)&:& M=6046.1~{\rm MeV}\, ,
\\ \nonumber \Omega^*_b(1/2^+)&:& M=6063~{\rm MeV}\, .
\end{eqnarray}

\item For the light pseudoscalar and vector mesons, we use
\begin{eqnarray}
\nonumber \pi(0^-)&:& m=138.04~{\rm MeV}\, ,
\\ \nonumber K(0^-) &:& m=495.65 ~{\rm MeV}\, ,
\\ \nonumber \rho(1^-)&:& m=775.21~{\rm MeV}\, ,
\\ \nonumber     &&\Gamma=148.2~{\rm MeV}\, ,
\\ \nonumber   &&g_{\rho\pi\pi}=5.94~{\rm GeV}^{-2}\, ,
\\ \nonumber K^{*}(1^-)&:& m=893.57 ~{\rm MeV}\, ,
\\ \nonumber &&\Gamma=49.1~{\rm MeV}\, ,
\\ \nonumber &&g_{K^* K\pi}=3.20~{\rm GeV}^{-2}\, .
\end{eqnarray}

\end{itemize}

\begin{table*}[ht]
\renewcommand{\arraystretch}{1.6}
\caption{Mass spectrum of the $P$-wave bottom baryons belonging to the  $SU(3)$ flavor antitriplet representation, calculated through the QCD sum rule method within the framework of heavy quark effective theory. Decay constants in the last column satisfy $f_{\Xi_b^0} = f_{\Xi_b^-}$. The table is taken from Ref.~\cite{Tan:2023opd}.}
\label{tabmass}
\begin{tabular}{ c |c | c | c | c | c c | c | c}
\hline\hline
\multirow{2}{*}{Multipluts}&\multirow{2}{*}{~B~} & $\omega_c$ & ~~~Working region~~~ & ~~~~~$\overline{\Lambda}$~~~~~ & ~~~Baryon~~~ & ~~~~Mass~~~~~ & Difference & Decay constant
\\                                               & & (GeV)      & (GeV)                & (GeV)                              & ($j^P$)       & (GeV)      & (MeV)        & (GeV$^{4}$)
\\ \hline\hline
 \multirow{4}{*}{$[\mathbf{\bar 3}_F,1,0,\lambda]$}
 &\multirow{2}{*}{$\Lambda_b$} & \multirow{2}{*}{$2.19\pm0.10$} & \multirow{2}{*}{$0.26< T < 0.40$} & \multirow{2}{*}{$1.53 ^{+0.14}_{-0.09}$} & $\Lambda_b(1/2^-)$ & $5.91^{+0.17}_{-0.13}$ & \multirow{2}{*}{$4 \pm 2$} & $0.081^{+0.020}_{-0.018}~(\Lambda^0_b(1/2^-))$
\\ \cline{6-7}\cline{9-9}
 & & & & & $\Lambda_b(3/2^-)$ & $5.91^{+0.17}_{-0.13}$ & &$0.038 ^{+0.009}_{-0.008}~(\Lambda^0_b(3/2^-))$
\\ \cline{2-9}
 &\multirow{2}{*}{$\Xi_b$} & \multirow{2}{*}{$2.36\pm0.10$} & \multirow{2}{*}{$0.25< T < 0.43$} & \multirow{2}{*}{$1.68^{+0.17}_{-0.07}$} & $\Xi_b(1/2^-)$ & $6.10^{+0.20}_{-0.10}$ & \multirow{2}{*}{$4 \pm 2$}
 &{ $0.110^{+0.030}_{-0.016}~(\Xi_b^-(1/2^-))$}
\\ \cline{6-7}\cline{9-9}
 & & & & & $\Xi_b(3/2^-)$ & $6.10^{+0.20}_{-0.10}$ & &{ $0.052 ^{+0.014}_{-0.008}~(\Xi_b^-(3/2^-))$}
\\ \hline
\multirow{2}{*}{$[\mathbf{\bar 3}_F,0,1,\rho]$}
&{$\Lambda_b$}&{$2.33\pm0.10$}&{$0.35< T < 0.44$} &{$1.56^{+0.14}_{-0.16}$}&
$\Lambda_b(1/2^-)$ & $5.92^{+0.17}_{-0.19}$ &{-} & $0.059 ^{+0.013}_{-0.013}~(\Lambda^0_b(1/2^-))$
\\ \cline{2-9}
&{$\Xi_b$} &{$2.29\pm0.10$}&{$ T=0.45,~PC=34\%$} &{$1.72 ^{+0.07}_{-0.07}$}&
$\Xi_b(1/2^-)$ & $6.10^{+0.08}_{-0.08}$ &{-} &{ $0.078 ^{+0.011}_{-0.009}
~(\Xi^-_b(1/2^-))$}
\\ \cline{1-9}
\multirow{4}{*}{$[\mathbf{\bar 3}_F,1,1,\rho]$}
&\multirow{2}{*}{$\Lambda_b$} & \multirow{2}{*}{$2.17\pm0.10$} & \multirow{2}{*}{$0.26< T < 0.39$} & \multirow{2}{*}{$1.60 ^{+0.11}_{-0.08}$} & $\Lambda_b(1/2^-)$ & $5.92^{+0.13}_{-0.10}$ & \multirow{2}{*}{$8\pm 3$} & $0.169 ^{+0.036}_{-0.027}~(\Lambda^0_b(1/2^-))$
\\ \cline{6-7}\cline{9-9}
 & & & & & $\Lambda_b(3/2^-)$ & $5.92^{+0.13}_{-0.10}$ & &$0.080 ^{+0.017}_{-0.013}~(\Lambda^0_b(3/2^-))$
\\ \cline{2-9}
 &\multirow{2}{*}{$\Xi_b$}& \multirow{2}{*}{$2.32\pm0.10$} & \multirow{2}{*}{$0.27< T < 0.41$} & \multirow{2}{*}{$1.75 ^{+0.11}_{-0.09}$} & $\Xi_b(1/2^-)$ & $6.09^{+0.13}_{-0.12}$ & \multirow{2}{*}{$7 \pm 3$} & { $0.222 ^{+0.069}_{-0.039}~(\Xi_b^-(1/2^-))$}
\\ \cline{6-7}\cline{9-9}
 & & & & & $\Xi_b(3/2^-)$ & $6.09^{+0.13}_{-0.12}$ & &{ $0.105 ^{+0.033}_{-0.018}~(\Xi_b^-(3/2^-))$}
\\ \hline
 \multirow{4}{*}{$[\mathbf{\bar 3}_F,2,1,\rho]$}
 &\multirow{2}{*}{$\Lambda_b$} &\multirow{2}{*}{ $2.17\pm0.10$ }& \multirow{2}{*}{$0.30< T < 0.37$} &\multirow{2}{*}{ $1.64^{+0.11}_{-0.11}$} & $\Lambda_b(3/2^-)$ & $5.93 ^{+0.13}_{-0.13}$ &\multirow{2}{*}{ $17\pm8$}& $0.136 ^{+0.031}_{-0.029}~(\Lambda^0_b(1/2^-))$
\\ \cline{6-7}\cline{9-9}
& & & & &$\Lambda_b(5/2^-)$ & $5.94^{+0.13}_{-0.13}$ & &$0.058 ^{+0.013}_{-0.012}~(\Lambda^0_b(5/2^-))$
\\ \cline{2-9}
 & \multirow{2}{*}{$\Xi_b$} & \multirow{2}{*}{$2.31\pm0.10$} & \multirow{2}{*}{$0.31< T < 0.40$} & \multirow{2}{*}{$1.80 ^{+0.09}_{-0.10}$} & $\Xi_b(3/2^-)$ & $6.10^{+0.15}_{-0.10}$ & \multirow{2}{*}{$14\pm 7$} & { $0.184 ^{+0.038}_{-0.036}~(\Xi^-_b(3/2^-))$}
 \\ \cline{6-7}\cline{9-9}
& & & & & $\Xi_b(5/2^-)$ & $6.11 ^{+0.15}_{-0.10}$ & &{ $0.078 ^{+0.016}_{-0.015}~(\Xi_b^-(5/2^-))$}
\\ \hline \hline
\end{tabular}
\end{table*}

\section{Strong decay properties}
\label{sec:decay}

In this section, we shall investigate strong decay properties of the $P$-wave single bottom baryons belonging to the $SU(3)$ flavor antitriplet representation via the light-cone sum rule method within the framework of heavy quark effective theory. We shall investigate their $S/D$-wave decays into ground-state bottom baryons with light pseudoscalar/vector mesons:
\begin{widetext}
\begin{eqnarray}
&(a1)& {\bf \Gamma\Big[} \Lambda_b[1/2^-,1P] \rightarrow \Lambda_b + \pi {\Big]}
=  {\bf \Gamma\Big[}\Lambda_b^0[1/2^-,1P] \rightarrow \Lambda_b^0 +\pi^0 {\Big]} \, ,
\\ &(a2)& {\bf \Gamma\Big[} \Lambda_b[1/2^-,1P] \rightarrow \Sigma_b + \pi {\Big]}
=3\times {\bf \Gamma\Big[} \Lambda_b^0[1/2^-,1P] \rightarrow \Sigma_b^{+}+\pi^-\to\Lambda_b^0\pi^++\pi^- {\Big]} \, ,
\\ &(a3)& {\bf \Gamma\Big[} \Lambda_b[1/2^-,1P] \rightarrow \Sigma_b^* + \pi\rightarrow\Lambda_b+\pi+\pi {\Big]
= 3 \times {\bf \Gamma \Big[}\Lambda_b^0[1/2^-,1P] \rightarrow \Sigma_b^{*+}+\pi^-\rightarrow \Lambda_b^0+\pi^++\pi^-{\Big ]}} \, ,
\\ &(a4)& {\bf\Gamma\Big[} \Lambda_b[1/2^-,1P] \rightarrow \Lambda_b + \rho \rightarrow\Lambda_b+\pi+\pi{\Big ]}
= {\bf \Gamma\Big[} \Lambda_b^0[1/2^-,1P] \rightarrow \Lambda_b^0 +\pi^++ \pi^- {\Big ]} \, ,
\\ &(a5)& { \bf\Gamma\Big[}\Lambda_b[1/2^-,1P] \rightarrow \Sigma_b + \rho\rightarrow\Sigma_b+\pi+\pi{\Big ]}
= 3 \times { \bf\Gamma\Big[}\Lambda_b^0[1/2^-,1P] \rightarrow \Sigma_b^0 +\pi^++ \pi^-{\Big ]} \, ,
\\ &(a6)&{\bf \Gamma\Big[}\Lambda_b[1/2^-,1P] \rightarrow \Sigma_b^* + \rho\rightarrow\Sigma_b^*+\pi+\pi{\Big ]}
= 3 \times { \bf\Gamma\Big[}\Lambda_b^0[1/2^-,1P] \rightarrow \Sigma_b^{*0} +\pi^++ \pi^-{\Big ]} \, ,
\\ &(b1)& {\bf \Gamma\Big[}\Lambda_b[3/2^-,1P] \rightarrow \Lambda_b + \pi{\Big ]}
= {\bf \Gamma\Big[}\Lambda_b^0[3/2^-] \rightarrow \Lambda_b^0 +\pi^0{\Big ]} \, ,
\\ &(b2)&{\bf \Gamma\Big[}\Lambda_b[3/2^-,1P] \rightarrow \Sigma_b + \pi\rightarrow\Lambda_b+\pi+\pi{\Big ]}
= 3 \times {\bf \Gamma\Big[}\Lambda_b^0[3/2^-,1P] \rightarrow \Sigma_b^{+} +\pi^-\rightarrow\Lambda_b^0\pi^++\pi^-{\Big ]} \, ,
\\ &(b3)&{\bf \Gamma\Big[}\Lambda_b[3/2^-,1P] \rightarrow \Sigma_b^* + \pi\rightarrow\Lambda_b+\pi+\pi{\Big ]}
= 3 \times {\bf \Gamma\Big[}\Lambda_b^0[3/2^-,1P] \rightarrow \Sigma_b^{*+} +\pi^-\rightarrow\Lambda_b^0\pi^++\pi^-{\Big ]} \, ,
\\ &(b4)&{\bf \Gamma\Big[} \Lambda_b[3/2^-,1P] \rightarrow \Lambda_b + \rho \rightarrow\Lambda_b+\pi+\pi{\Big ]}
= { \bf\Gamma\Big[}\Lambda_b^0[3/2^-,1P] \rightarrow \Lambda_b^0 +\pi^++ \pi^- {\Big ]} \, ,
\\ &(b5)& { \bf\Gamma\Big[}\Lambda_b[3/2^-,1P] \rightarrow \Sigma_b + \rho\rightarrow\Sigma_b+\pi+\pi{\Big ]}
= 3 \times { \bf\Gamma\Big[}\Lambda_b^0[3/2^-,1P] \rightarrow \Sigma_b^0 \pi^++ \pi^-{\Big ]} \, ,
\\&(b6)& { \bf\Gamma\Big[}\Lambda_b[3/2^-,1P] \rightarrow \Sigma_b^* + \rho\rightarrow\Sigma_c^*+\pi+\pi {\Big ]}
= 3 \times {\bf \Gamma\Big[}\Lambda_b^0[3/2^-,1P] \rightarrow \Sigma_b^{*+} + \pi^++\pi^- {\Big ]} \, ,
\\ &(c1)& {\bf \Gamma\Big[}\Xi_b[1/2^-,1P] \rightarrow \Lambda_b + \bar K{\Big ]}
= {\bf \Gamma\Big[}\Xi_b^-[1/2^-,1P] \rightarrow\Lambda_b^0 +K^-{\Big ]} \, ,
\\ &(c2)&{\bf \Gamma\Big[}\Xi_b[1/2^-,1P] \rightarrow\Xi_b + \pi{\Big ]}
= {3\over2} \times {\bf \Gamma\Big[}\Xi_b^-[1/2^-,1P] \rightarrow \Xi_b^0 +\pi^-{\Big ]} \, ,
\\ &(c3)&{\bf \Gamma\Big[}\Xi_b[1/2^-,1P] \rightarrow\Sigma_b + \bar K{\Big ]}
= 3 \times {\bf \Gamma\Big[}\Xi_b^-[1/2^-,1P] \rightarrow \Sigma_b^0 +K^-{\Big ]} \, ,
\\ &(c4)& { \bf\Gamma\Big[}\Xi_b[1/2^-,1P] \rightarrow \Xi_b^{\prime}+\pi {\Big]}
= {3\over2}\times{\bf \Gamma\Big[}\Xi_b^-[1/2^-,1P]\rightarrow\Xi_b^{\prime0}+\pi^-{\Big]} \, ,
\\ &(c5)& {\bf \Gamma\Big[}\Xi_b[1/2^-,1P] \rightarrow \Sigma_b^* + K{\Big ]}
= 3 \times {\bf \Gamma\Big[}\Xi_b^-[1/2^-,1P] \rightarrow \Sigma_b^{*0} + K^-{\Big ]} \, ,
\\ &(c6)& {\bf \Gamma\Big[}\Xi_b[1/2^-,1P] \rightarrow \Xi_b^* + \pi\rightarrow \Xi_b+\pi+\pi {\Big ]}
= {9 \over 2} \times {\bf \Gamma\Big[}\Xi_b^{-}[1/2^-,1P] \rightarrow\Xi_b^{*0} + \pi^-\rightarrow\Xi_b^0+\pi^0+\pi^-{\Big]} \, ,
\\ &(c7)& {\bf \Gamma\Big[}\Xi_b[1/2^-,1P] \rightarrow \Lambda_b + \bar{K}^*\rightarrow\Lambda_b+\bar K+\pi{\Big ]}
=3\times  {\bf \Gamma\Big[}\Xi_b^-[1/2^-,1P] \rightarrow \Lambda_b^0 +  K^- +\pi^0{\Big ]} \, ,
\\ &(c8)& {\bf \Gamma\Big[}\Xi_b[1/2^-,1P] \rightarrow\Xi_b + \rho\rightarrow\Xi_c+\pi+\pi{\Big ]}
= {3\over2} \times {\bf \Gamma\Big[}\Xi_b^-[1/2^-,1P] \rightarrow \Xi_b^0 + \pi^0+\pi^-{\Big ]} \, ,
\\ &(c9)& {\bf \Gamma\Big[}\Xi_b[1/2^-,1P] \rightarrow \Sigma_b^* + \bar K^*\rightarrow\Sigma_b^{*}+\bar K+\pi{\Big ]}
= 9 \times {\bf \Gamma\Big[}\Xi_b^-[1/2^-,1P] \rightarrow \Sigma_b^{*0} + K^-+ \pi^0{\Big ]} \, ,
\\ &(c10)& {\bf \Gamma\Big[}\Xi_b[1/2^-,1P] \rightarrow \Xi_b^{*}+ \rho\rightarrow\Xi_b^{*}+\pi+\pi {\Big ]}
={3\over2} \times {\bf \Gamma\Big[}\Xi_b^-[1/2^-,1P] \rightarrow \Xi_b^{*0} + \pi^0+\pi^- {\Big ]}\, ,
\\ &(d1)& {\bf \Gamma\Big[}\Xi_b[3/2^-,1P] \rightarrow \Lambda_b + \bar K{\Big ]}
= {\bf \Gamma\Big[}\Xi_b^-[3/2^-,1P] \rightarrow\Lambda_b^0 +K^-{\Big ]} \, ,
\\ &(d2)&{\bf \Gamma\Big[}\Xi_b[3/2^-,1P] \rightarrow\Xi_b + \pi{\Big ]}
= {3\over2} \times {\bf \Gamma\Big[}\Xi_b^-[3/2^-,1P] \rightarrow \Xi_b^0 +\pi^-{\Big ]} \, ,
\\ &(d3)&{\bf \Gamma\Big[}\Xi_b[3/2^-,1P] \rightarrow \Sigma_b + \bar K{\Big ]}
= 3 \times {\bf \Gamma\Big[}\Xi_b^-[3/2^-,1P] \rightarrow \Sigma_b^0 +K^-{\Big ]} \, ,
\\ &(d4)& { \bf\Gamma\Big[}\Xi_b[3/2^-,1P] \rightarrow \Xi_b^{\prime}+\pi {\Big]}
= {3\over2}\times{\bf \Gamma\Big[}\Xi_b^-[3/2^-,1P]\rightarrow\Xi_b^{\prime0}+\pi^-{\Big]} \, ,
\\ &(d5)& {\bf \Gamma\Big[}\Xi_b[3/2^-,1P] \rightarrow \Sigma_b^* + \bar K{\Big ]}
= 3 \times {\bf \Gamma\Big[}\Xi_b^-[3/2^-,1P] \rightarrow \Sigma_b^{*0} + K^-{\Big ]} \, ,
\\ &(d6)& {\bf \Gamma\Big[}\Xi_b[3/2^-,1P] \rightarrow \Xi_b^* + \pi\rightarrow \Xi_b+\pi+\pi {\Big ]}
= {9 \over 2} \times {\bf \Gamma\Big[}\Xi_b^{*-}[3/2^-,1P] \rightarrow\Xi_b^{*0} + \pi^-\rightarrow\Xi_b^0+\pi^0+\pi^-{\Big]} \, ,
\\ &(d7)& {\bf \Gamma\Big[}\Xi_b[3/2^-,1P] \rightarrow \Lambda_b + \bar{K}^*\rightarrow\Lambda_b+\bar K+\pi{\Big ]}
=3\times  {\bf \Gamma\Big[}\Xi_b^-[3/2^-,1P] \rightarrow \Lambda_b^0 +  K^- +\pi^0{\Big ]} \, ,
\\ &(d8)& {\bf \Gamma\Big[}\Xi_b[3/2^-,1P] \rightarrow \Xi_b + \rho\rightarrow\Xi_b+\pi+\pi{\Big ]}
= {3\over2} \times {\bf \Gamma\Big[}\Xi_b^-[3/2^-,1P] \rightarrow \Xi_b^0 + \pi^0+\pi^-{\Big ]} \, ,
\\ &(d9)& {\bf \Gamma\Big[}\Xi_b[3/2^-,1P] \rightarrow \Sigma_b^* + \bar K^*\rightarrow\Sigma_b^{*0}+\bar K+\pi{\Big ]}
= 9 \times {\bf \Gamma\Big[}\Xi_b^-[3/2^-,1P] \rightarrow \Sigma_b^{*0} + K^-+ \pi^0{\Big ]} \, ,
\\ &(d10)& {\bf \Gamma\Big[} \Xi_b[3/2^-,1P] \rightarrow \Xi_b^{*}+ \rho\rightarrow\Xi_b^{*}+\pi+\pi {\Big ]}
={3\over2} \times {\bf \Gamma\Big[}\Xi_b^-[3/2^-,1P] \rightarrow \Xi_b^{*0} + \pi^0+\pi^- {\Big ]}\, .
\label{eq:couple}
\end{eqnarray}
\end{widetext}
Their partial decay widths can be evaluated through the following Lagrangians:
\begin{eqnarray}
&&\mathcal{L}^S_{X_b({1/2}^-) \rightarrow Y_b({1/2}^+) P}
\\ \nonumber&& ~~~~~~~~~~~~\, = g^S {\bar X_b}(1/2^-) Y_b(1/2^+) P \, ,
\\ &&\mathcal{L}^S_{X_b({3/2}^-) \rightarrow Y_b({3/2}^+) P}
\\ \nonumber&& ~~~~~~~~~~~~\, = g^S {\bar X_{b\mu}}(3/2^-)Y_b^{\mu}(3/2^+) P \, ,
\\ &&\mathcal{L}^S_{X_b({1/2}^-) \rightarrow Y_b({1/2}^+) V}
\\ \nonumber&& ~~~~~~~~~~~~\, = g^S {\bar X_b}(1/2^-) \gamma_\mu \gamma_5 Y_b(1/2^+) V^\mu \, ,
\\ &&\mathcal{L}^S_{X_b({1/2}^-) \rightarrow Y_b({3/2}^+) V}
\\ \nonumber&& ~~~~~~~~~~~~\, = g^S {\bar X_{b}}(1/2^-) Y_{b}^{\mu}(3/2^+) V_\mu \, ,
\\ &&\mathcal{L}^S_{X_b({3/2}^-) \rightarrow Y_b({1/2}^+) V}
\\ \nonumber&& ~~~~~~~~~~~~\, = g^S {\bar X_{b}^{\mu}}(3/2^-) Y_{b}(1/2^+) V_\mu \, ,
\\ &&\mathcal{L}^S_{X_b({3/2}^-) \rightarrow Y_b({3/2}^+) V}
\\ \nonumber&& ~~~~~~~~~~~~\, = g^S {\bar X_b}^{\nu}(3/2^-) \gamma_\mu \gamma_5 Y_{b\nu}(3/2^+) V^\mu \, ,
\\&& \mathcal{L}^S_{X_b({5/2}^-) \rightarrow Y_b({3/2}^+) V}
\\ \nonumber&& ~~~~~~~~~~~~\, = g^S {\bar X_{b}^{\mu\nu}}(5/2^-) Y_{b\mu}(3/2^+) V_\nu
\\ \nonumber &&~~~~~~~~~~~~\, + g^S {\bar X_{b}^{\nu\mu}}(5/2^-) Y_{b\mu}(3/2^+) V_\nu \, ,
\\ && \mathcal{L}^D_{X_b({1/2}^-) \rightarrow Y_b({3/2}^+) P}
\\ \nonumber && ~~~~~~~~~~~\, = g^D {\bar X_b}(1/2^-) \gamma_\mu \gamma_5 Y_{b\nu}(3/2^+) \partial^{\mu} \partial^{\nu}P \, ,
\\ && \mathcal{L}^D_{X_b({3/2}^-) \rightarrow Y_b({1/2}^+) P}
\\ \nonumber && ~~~~~~~~~~~\, = g^D {\bar X_{b\mu}}(3/2^-) \gamma_\nu \gamma_5 Y_{b}(1/2^+) \partial^{\mu} \partial^{\nu}P \, ,
\label{eq:lagrangians}
\\ && \mathcal{L}^D_{X_b({3/2}^-) \rightarrow Y_b({3/2}^+) P}
\\ \nonumber && ~~~~~~~~~~~\, = g^D {\bar X_{b\mu}}(3/2^-) Y_{b\nu}(3/2^+) \partial^{\mu} \partial^{\nu}P \, ,
\\ && \mathcal{L}^D_{X_b({5/2}^-) \rightarrow Y_b({1/2}^+) P}
\\ \nonumber && ~~~~~~~~~~~\, = g^D {\bar X_{b\mu\nu}}(5/2^-) Y_{b}(1/2^+) \partial^{\mu} \partial^{\nu}P \, ,
\\ && \mathcal{L}^D_{X_b({5/2}^-) \rightarrow Y_b({3/2}^+) P}
\\ \nonumber && ~~~~~~~~~~~\, = g^D {\bar X_{b\mu\nu}}(5/2^-) \gamma_\rho \gamma_5 Y_{b}^{\mu}(3/2^+) \partial^{\nu} \partial^{\rho}P
\\ \nonumber && ~~~~~~~~~~~\, + g^D {\bar X_{b\mu\nu}}(5/2^-) \gamma_\rho \gamma_5 Y_{b}^{\nu}(3/2^+) \partial^{\mu} \partial^{\rho}P \, .
\end{eqnarray}
In the above expressions, $S$ and $D$ denote the $S$- and $D$-wave decays, respectively; the fields $X_b^{(\mu\nu)}$ and $Y_b^{(\mu)}$ denote the $P$-wave and ground-state bottom baryons, respectively.

As an example, we select the $P$-wave bottom baryon $\Lambda_b^0({3/2}^-)$ belonging to the $[\mathbf{\bar 3}_F, 1, 1, \rho]$ doublet, and calculate its $D$-wave decay into $\Sigma_b^{+}\pi^-$. To achieve this, we consider the two-point correlation function:

\begin{eqnarray}
&& \Pi^{\alpha}(\omega, \omega^\prime)
\\ \nonumber &=& \int d^4 x~e^{-i k \cdot x}~\langle 0 | J^\alpha_{3/2,-,\Lambda_b^0,1,1,\rho}(0) \bar J_{\Sigma_b^{+}}(x) | \pi^-(q) \rangle
\\ \nonumber
 \\ \nonumber &=& {1+v\!\!\!\slash\over2} G^{\alpha}_{\Lambda_b^0[{3\over2}^-] \rightarrow   \Sigma_b^{+}\pi^-} (\omega, \omega^\prime) \, ,
\end{eqnarray}
where $k^\prime = k + q$, $\omega = v \cdot k$, and $\omega^\prime = v \cdot k^\prime$.

At the hadron level, we write
$G^{\alpha}_{\Lambda_b^0[{3\over2}^-] \rightarrow \Sigma_b^{+}\pi^-}$ as
\begin{eqnarray}
&&G^{\alpha}_{\Lambda_b^0[{3\over2}^-] \rightarrow \Sigma_b^{+}\pi^-} (\omega, \omega^\prime)
\label{G0C}
\\ \nonumber
&=& g^D_{\Lambda_b^0[{3\over2}^-] \rightarrow \Sigma_b^{+}\pi^-} {  f_{\Lambda_b^0[{3\over2}^-]} f_{\Sigma_b^{+}} \over (\bar \Lambda_{\Lambda_b^0[{3\over2}^-]} - \omega^\prime) (\bar \Lambda_{\Sigma_b^{+}} - \omega)} \gamma_{\alpha}\gamma_5+ \cdots \, ,
\end{eqnarray}
where $\cdots$ contains other possible amplitudes.

\begin{widetext}
At the quark-gluon level, we calculate $G^\alpha_{\Lambda_b^0[{3\over2}^-] \rightarrow \Sigma_b^{+}\pi^-}$ using the method of operator product expansion (OPE):
\begin{eqnarray}
\label{eq:g1}
&& G^{\alpha}_{\Lambda_b^0[{3\over2}^-] \rightarrow \Sigma_b^{+}\pi^-} (\omega, \omega^\prime)
\\ \nonumber &=& \int_0^\infty dt \int_0^1 du e^{i (1-u) \omega^\prime t} e^{i u \omega t}\times4 \Bigg(-\frac{{{\rm{}}{f_\pi u_0}}}{{4{\pi ^2t^3}}}{\phi _{2;\pi }}+\frac{ f_{\pi}{u_0}}{64 \pi^2t}\phi_{4;\pi}(u_0)+\frac{ f_{\pi}m^2_{\pi}u_0t}{144(m_{u})}\langle\bar q q\rangle\phi _3^\sigma(u_0)
\\ \nonumber
&-&\frac{{{f_\pi }m_\pi ^2{u_0t^3}}}{{2304({m_u})}}\langle {g_s}\sigma Gs\rangle \phi _3^\sigma ({u_0}) \Bigg)\times\gamma_{\alpha}\gamma_5 +\int_0^\infty dt \int_0^1 du \int \mathcal{D} \underline{\alpha} e^{i \omega^{\prime} t(\alpha_2 + u \alpha_3)} e^{i \omega t(1 - \alpha_2 - u \alpha_3)}\times \frac{1}{2}\Bigg( \frac{{{f_\pi }{\Phi _{4;\pi }}(\underline\alpha){\alpha _3}{u_0}}}{{24{\pi ^2t}}}
\\ \nonumber
&+& \frac{{{f_\pi }{\Phi _{4;\pi }}(\underline\alpha){\alpha _2}{u_0}}}{{24{\pi ^2t}}}
+ \frac{{{f_\pi }{\Phi _{4;\pi }}(\underline\alpha){\alpha _3}{u_0}}}{{48{\pi ^2t}}}+\frac{{{f_\pi }{\widetilde\Phi _{4;\pi }}(\underline\alpha){\alpha _3}{u_0}}}{{48{\pi ^2t}}}- \frac{{{f_\pi }{\Phi _{4;\pi }}(\underline\alpha){u_0}}}{{24{\pi ^2t}}} + \frac{{{f_\pi }{\Phi _{4;\pi }}(\underline\alpha){\alpha _2}}}{{48{\pi ^2t}}}
+ \frac{{{f_\pi }{\widetilde\Phi _{4;\pi }}(\underline\alpha){\alpha _2}}}{{48{\pi ^2t}}}
\\ \nonumber
&-&\frac{{{f_\pi }{\Phi _{4;\pi }}(\underline\alpha)}}{{48{\pi ^2t}}}
 -\frac{{{f_\pi }{\widetilde\Phi _{4;\pi }}(\underline\alpha)}}{{48{\pi ^2t}}})
-\frac{{{f_\pi }{\Phi _{4;\pi }}(\underline\alpha)u_0}}{{12{\pi ^2t^2(v\cdot q)}}} + \frac{{{f_\pi }\widetilde\Phi _{4;\pi }(\underline\alpha){u_0}}}{{4{\pi ^2t^2(v\cdot q)}}} + \frac{{{f_\pi }\Psi_{4;\pi}(\underline\alpha){u_0}}}{{24{\pi ^2t^2(v\cdot q)}}}
  + \frac{{{f_\pi }\widetilde\Psi _{4;\pi }(\underline\alpha){u_0}}}{{8{\pi ^2t^2(v\cdot q)}}}
 + \frac{{{f_\pi }{\Phi_{4;\pi }}(\underline\alpha)}}{{8{\pi ^2t^2(v\cdot q)}}}
  \\  \nonumber
  &+& \frac{{{f_\pi }\widetilde\Phi _{4;\pi }}(\underline\alpha)}{{24{\pi ^2t^2(v\cdot q)}}}
+ \frac{{{f_\pi }{\Phi _{4;\pi }}(\underline\alpha)}}{{24{\pi ^2t^2(v\cdot q)}}}- \frac{{{f_\pi } \widetilde\Phi _{4;\pi }(\underline\alpha)}}{{24{\pi ^2t^2(v\cdot q)}}}\Bigg)\times\gamma_{\alpha}\gamma_5 \, .
\end{eqnarray}
Where $\int\mathcal{D}{\underline{\alpha}}=\int_0^1 d\alpha_1d\alpha_2d\alpha_3\delta(1-\alpha_1-\alpha_2-\alpha_3)$.
 Then we perform the Borel transformation to both Eq.~(\ref{G0C}) at the hadron level and Eq.~(\ref{eq:g1}) at the quark-gluon level:
\begin{eqnarray}
\label{eq:g}
&& g^D_{\Lambda_b^0[{3\over2}^-] \rightarrow\Sigma_b^{+}\pi^-} f_{\Lambda_b^0[{3\over2}^-]}
f_{\Sigma_b^{+}} e^{- {\bar \Lambda_{\Lambda_b^0[{3\over2}^-]} \over T_1}} e^{ - {\bar \Lambda_{\Sigma_b^{+}} \over T_2}}
\\ \nonumber &=& 4  \Bigg(-\frac{{{\rm{}}{f_\pi u_0 }}}{{4{\pi ^2}}}{T^4}{f_3}(\frac{{{\omega _c}}}{T}){\phi _{2;\pi }}({u_0})|_{u_0 = \frac{1}{2}}+\frac{ f_{\pi}{u_0}}{64 \pi^2}T^2f_1(\frac{\omega_c}{T})\phi_{4;\pi}(u_0)u_0|_{u_0 = \frac{1}{2}}+\frac{ f_{\pi}m^2_{\pi}u_0}{144(m_{u})}\langle\bar q q\rangle\phi _3^\sigma(u_0)|_{u_0 = \frac{1}{2}}
\\ \nonumber
&-&\frac{{{f_\pi }m_\pi ^2{u_0}}}{{2304({m_u})}}{T^{-2}}\langle {g_s}\sigma Gs\rangle \phi _3^\sigma ({u_0})|{_{{u_0} = \frac{1}{2}}} \Bigg) + \frac{1}{2} \times ({f_1}(\frac{{{\omega _c}}}{T})\int_0^{\frac{1}{2}} d {\alpha _2}\int_{\frac{1}{2} - {\alpha_2}}^{1 - {\alpha _2}}d {\alpha _3}  \frac{{{{T}^2}}}{{\alpha_3}}\Bigg(\frac{{{f_\pi }{\Phi _{4;\pi }}(\underline\alpha){\alpha _3}{u_0}}}{{24{\pi ^2}}} + \frac{{{f_\pi }{\Phi _{4;\pi }}(\underline\alpha){\alpha _2}{u_0}}}{{24{\pi ^2}}}
\\ \nonumber
 &+& \frac{{{f_\pi }{\Phi _{4;\pi }}(\underline\alpha){\alpha _3}{u_0}}}{{48{\pi ^2}}}+\frac{{{f_\pi }{\widetilde\Phi _{4;\pi }}(\underline\alpha){\alpha _3}{u_0}}}{{48{\pi ^2}}}- \frac{{{f_\pi }{\Phi _{4;\pi }}(\underline\alpha){u_0}}}{{24{\pi ^2}}} + \frac{{{f_\pi }{\Phi _{4;\pi }}(\underline\alpha){\alpha _2}}}{{48{\pi ^2}}}+ \frac{{{f_\pi }{\widetilde\Phi _{4;\pi }}(\underline\alpha){\alpha _2}}}{{48{\pi ^2}}}-\frac{{{f_\pi }{\Phi _{4;\pi }}(\underline\alpha)}}{{48{\pi ^2}}}
 -\frac{{{f_\pi }{\widetilde\Phi _{4;\pi }}(\underline\alpha)}}{{48{\pi ^2}}}\Bigg)
 \\ \nonumber
 &-&\frac{1}{2}\times{f_1}(\frac{{{\omega _c}}}{T})\int_0^{\frac{1}{2}} d {\alpha _2}\int_{\frac{1}{2} - {\alpha _2}}^{1 - {\alpha _2}} d {\alpha _4}\int_{0}^{\alpha_4}d{\alpha_3}\frac{{{u_0}{T^2}}}{{\alpha_4}}\Bigg(\frac{{{f_\pi }{\Phi _{4;\pi }}(\underline\alpha)u_0}}{{12{\pi ^2}}} + \frac{{{f_\pi }\widetilde\Phi _{4;\pi }(\underline\alpha){u_0}}}{{4{\pi ^2}}} + \frac{{{f_\pi }\Psi_{4;\pi}(\underline\alpha){u_0}}}{{24{\pi ^2}}}
  + \frac{{{f_\pi }\widetilde\Psi _{4;\pi }(\underline\alpha){u_0}}}{{8{\pi ^2}}}
 \\  \nonumber
 &+& \frac{{{f_\pi }{\Phi_{4;\pi }}(\underline\alpha)}}{{8{\pi ^2}}}+ \frac{{{f_\pi }\widetilde\Phi _{4;\pi }}(\underline\alpha)}{{24{\pi ^2}}}+ \frac{{{f_\pi }{\Phi _{4;\pi }}(\underline\alpha)}}{{24{\pi ^2}}}- \frac{{{f_\pi } \widetilde\Phi _{4};\pi }(\underline\alpha)}{{24{\pi ^2}}}\Bigg) \, .
\end{eqnarray}
\end{widetext}
In the above expressions, 
 $\alpha = \{ \alpha_1, \alpha_2, \alpha_3 \}$ and $\int \mathcal{D}\underline{\alpha} = \int_0^1 d\alpha_1 \int_0^1 d\alpha_2 \int_0^1 d\alpha_3$; the functions $\Phi_{\cdots}(\underline\alpha)$ in Eq.~(\ref{eq:g1}) contain the constraint $\delta(\alpha_1 + \alpha_2 + \alpha_3 - 1)$, whereas this $\delta$ function is integrated out in Eq.~(\ref{eq:g}); $f_n(x) \equiv 1 - e^{-x} \sum_{k=0}^n {x^k \over k!}$; the parameters $\omega$ and $\omega^\prime$ are transformed to be $T_1$ and $T_2$, respectively; we select the symmetric point $T_1 = T_2 = 2T$ so that $u_0 = {T_1 \over T_1 + T_2} = {1\over2}$; we choose $\omega_c = 2.17$~GeV to be the averaged threshold value of the $\Lambda_b^0({3/2}^-)$ and $\Sigma_b^{+}(3/2^+)$ mass sum rules; we select $0.26~\rm{GeV}<T<0.39~\rm{GeV}$ to be the Borel window of the $\Lambda_b^0({3/2}^-)$ mass sum rule. The light-cone distribution amplitudes contained in the above sum rule expressions can be found in Refs.~\cite{Ball:1998je,Ball:2006wn,Ball:2004rg,Ball:1998kk,Ball:1998sk,Ball:1998ff,Ball:2007rt,Ball:2007zt}.

\begin{figure}[hbt]
\begin{center}
\scalebox{0.8}{\includegraphics{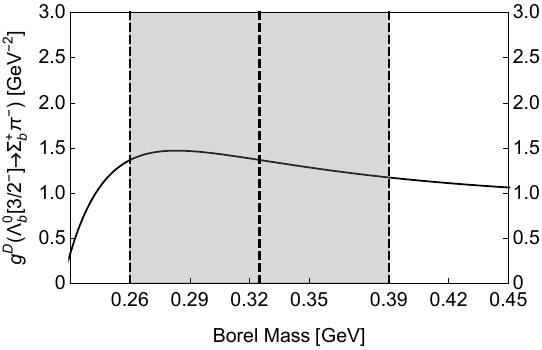}}
\caption{Coupling constant $g^D_{\Lambda^0_b[\frac{3}{2}^-]\to\Sigma_b^{+}\pi^-}$ with respect to the Borel mass $T$. Here the bottom baryon doublet $[\mathbf{\bar 3}_F, 1, 1, \rho]$ is investigated.
\label{fig:311R}}
\end{center}
\end{figure}

In the present work we use the following values for various quark and gluon parameters at the renormalization scale 1~GeV~\cite{Yang:1993bp,Hwang:1994vp,Ovchinnikov:1988gk,Narison:2002woh,Jamin:2002ev,Ioffe:2002be,Shifman:2001ck,Gimenez:2005nt}:
%
\begin{eqnarray}
\nonumber && \langle \bar qq \rangle = - (0.24 \pm 0.01)^3 \mbox{ GeV}^3 \, ,
\\ \nonumber && \langle \bar ss \rangle = (0.8\pm 0.1)\times \langle\bar qq \rangle \, ,
\\ && \langle g_s \bar q \sigma G q \rangle = M_0^2 \times \langle \bar qq \rangle\, ,
\label{eq:condensates}
\\ \nonumber && \langle g_s \bar s \sigma G s \rangle = M_0^2 \times \langle \bar ss \rangle\, ,
\\ \nonumber && M_0^2= 0.8 \mbox{ GeV}^2\, ,
\\ \nonumber && \langle g_s^2GG\rangle =(0.48\pm 0.14) \mbox{ GeV}^4\, .
\end{eqnarray}
As depicted in Fig.~\ref{fig:311R}, we extract the coupling constant from Eq.~(\ref{eq:g}) to be
\begin{eqnarray}
\nonumber g^D_{\Lambda_b^0[{3\over2}^-] \rightarrow \Sigma_b^{+}\pi^-}
&=&1.36~^{+0.11}_{-0.10}~^{+0.25}_{-0.15}~^{+0.28}_{-0.23}~^{+0.48}_{-0.49}{~\rm GeV}^{-2}
\\ \nonumber
 &=&1.36^{+0.94}_{-0.70}{~\rm GeV}^{-2} \, ,
\end{eqnarray}
where the uncertainties come from the Borel mass, parameters of $\Sigma_b^{+}(1/2^+)$, parameters of $\Lambda_b^0({3/2}^-)$, and various QCD parameters given in Eqs.~(\ref{eq:condensates}), respectively.

The $D$-wave decay widths of $P$-wave bottom baryons can be calculated through the Lagrangian given in Eq.~(\ref{eq:lagrangians}):
\begin{eqnarray}
&&\Gamma^D\left(\Lambda_b(3/2^-)\to\Sigma_b+\pi\right)\\
\nonumber &\equiv&3 \times\Gamma^D\left(\Lambda_b^0(3/2^-)\to\Sigma_b^{+}+\pi^-\right)\\
\nonumber &=& {3|\vec{p_2 }| g_{0\to 1+2}^2\over 32\pi m_0^2} \times \rm Tr[(p_0\!\!\!\slash+m_0)\gamma_5\gamma_{\beta1}(p_1\!\!\!\slash+m_1)\gamma_{\beta2}\gamma_5]
\\ \nonumber
&\times&({g_{{\alpha _1}{\alpha _2}}} - \frac{1}{3}{\gamma _{{\alpha _2}}}{\gamma _{{\alpha _1}}} - \frac{{{p_0^{{\alpha _2}}}{\gamma _{{\alpha _1}}} - {p_0^{{\alpha _1}}}{\gamma _{{\alpha _2}}}}}{{3m_0}} - \frac{{2{p_0^{{\alpha _2}}}{p_0^{{\alpha _1}}}}}{{3{m_0^2}}})
\\ \nonumber
&\times& p_2^{\alpha_1}p_2^{\alpha_2}p_2^{\beta_1}p_2^{\beta_2}\, ,
\end{eqnarray}
where $0$, $1$, and $2$ denote $\Lambda_b^0(3/2^-)$, $\Sigma_b^0(1/2^+)$, and $\pi^-$, respectively. Numerically, we obtain
\begin{eqnarray}
\Gamma^D_{\Lambda_b^0[{3\over2}^-] \rightarrow \Sigma_c\pi }&=& 1.67^{+3.08}_{-1.27}\times10^{-3}{\rm~MeV} \, .
\end{eqnarray}

Similarly, we study the four $SU(3)$  flavor antitriplet bottom baryon multiplets, $[\mathbf{\bar 3}_F, 1, 0, \lambda]$, $[\mathbf{\bar 3}_F, 0, 1, \rho]$, $[\mathbf{\bar 3}_F, 1, 1, \rho]$, and $[\mathbf{\bar 3}_F, 2, 1, \rho]$. We systematically study their $D$-wave decays into ground-state bottom baryons with light pseudoscalar mesons and $S$-wave decays into ground-state bottom baryons with light vector mesons. The relevant sum rule equations are listed in Appendix~\ref{sec:othersumrule}, and the obtained results are summarized in Table~\ref{tab:decayb3f}.

\begin{table*}[hbt]
\begin{center}
\renewcommand{\arraystretch}{1.4}
\caption{Decay properties of the $P$-wave bottom baryons belonging to the $SU(3)$ flavor antitriplet  representation. Radiative decay widths are given in the eighth column for the excited $\Lambda_b^0$ and $\Xi_b^0$ baryons~\cite{Luo:2024jov}.}
\setlength{\tabcolsep}{0.1mm}{
\begin{tabular}{c| c | c | c | c | c | c | c | c}
\hline\hline
\multirow{2}{*}{Multiplets}& ~B~ & ~~Mass~~ & Difference & \multirow{2}{*}{~~~~~Decay channels~~~~~}  & ~$g^{S/D}$~  & ~Decay width~ & ~Total width~ & \multirow{2}{*}{Candidate}
\\ & ($j^P$) & ({GeV})& ({MeV}) & & ($\rm GeV^{-2}$) & ({MeV}) & ({MeV})
\\ \hline\hline
\multirow{12}{*}{$[\mathbf{\bar 3}_F,1,0,\lambda]$}&\multirow{3}{*}{$\Lambda_b({1\over2}^-)$}&\multirow{3}{*}{$5.91^{+0.17}_{-0.13}$}
&\multirow{6}{*}{$4\pm2$}&$\Lambda_b({1\over2}^-)\to \Sigma_b\pi\to\Lambda_b\pi\pi$ &$g^S=17.98^{+10.06}_{-11.67}$&$0\sim32000$&\multirow{2}{*}{$\Gamma_{\rm EM}=0.45^{+5.90}_{-0.41}$}
&\multirow{3}{*}{--}
\\ \cline{5-7}
&&&&$\Lambda_b({1\over2}^-)\to \Sigma_b^*\pi\to\Lambda_b\pi\pi$&$g^D=3.28^{+1.75}_{-1.05}$&$0.0^{+2.6}_{-0.0}\times 10^{-5}$&\multirow{2}{*}{$\Gamma_{\rm Total}=0\sim32000$}&
\\ \cline{5-7}
&&&&$\Lambda_b({1\over2}^-)\to \Lambda_b\rho\to\Lambda_b\pi\pi$&$g^S=1.83^{+0.74}_{-0.52}$&$0.0^{+7.0}_{-0.0}$&&
\\ \cline{2-3} \cline{5-9}
&\multirow{3}{*}{$\Lambda_b({3\over2}^-)$}&\multirow{3}{*}{$5.91^{+0.17}_{-0.13}$}&
&\multirow{2}{*}{$\Lambda_b({3\over2}^-)\to\Sigma_b^*\pi\to\Lambda_b\pi\pi$}&$g^S=8.62^{+4.34}_{-4.81}$
&$0\sim6800$ &
\\&&&&&$g^D=2.32^{+1.22}_{-0.73}$&$0\sim20$&\multirow{1}{*}{$\Gamma_{\rm EM}=0.22^{+3.70}_{-0.22}$}&\multirow{2}{*}{--}
\\ \cline{5-7}
&&&&$\Lambda_b({3\over2}^-)\to\Sigma_b\pi$&$g^D=4.03^{+2.09}_{-1.40}$&$2.0^{+98.0}_{-~2.0}\times 10^{-2}$&\multirow{1}{*}{$\Gamma_{\rm Total}=0\sim6800$}&
\\ \cline{5-7}
&&&&$\Lambda_b({3\over2}^-)\to \Lambda_b\rho\to\Lambda\pi \pi$ &$g^S= 2.58^{+1.02}_{-0.71}$ &$0.0^{+0.5}_{-0.0}$ &&
\\ \cline{2-9}
&\multirow{3}{*}{$\Xi_b({1\over2}^-)$}&\multirow{3}{*}{$6.10^{+0.20}_{-0.10}$}
&\multirow{6}{*}{$4\pm2$}& $\Xi_b({1\over2}^-)\to \Xi_b^{\prime}\pi$&$g^S=12.05^{+6.72}_{-5.71}$&$2800^{+8200}_{-2800}$&\multirow{2}{*}{$\Gamma_{\rm EM}=1.8^{+17.0}_{-~1.8}$}&\multirow{3}{*}{--}
\\ \cline{5-7}
&&&&$\Xi_b({1\over2}^-)\to\Xi_b^*\pi\to\Xi_b\pi\pi$&$g^D=1.98^{+1.12}_{-0.67}$&$0.0^{+2.0}_{-0.0}\times 10^{-6}$&\multirow{2}{*}{$\Gamma_{\rm Total}=2800^{+8200}_{-2800}$}&
\\ \cline{5-7}
&&&&$\Xi_b({1\over2}^-)\to\Xi_b\rho\to\Xi_b\pi\pi$&$g^S=1.32^{+0.47}_{-0.37} $&$0.0^{+1.4}_{-0.0}$&
\\ \cline{2-3} \cline{5-9}
&\multirow{3}{*}{$\Xi_b({3\over2}^-)$}&\multirow{3}{*}{$6.10^{+0.20}_{-0.10}$}&
&$\Xi_b({3\over2}^-)\to\Xi_ b^{\prime}\pi$&$g^D=2.42^{+1.28}_{-0.83}$&$0.0^{+2.0}_{-0.0}$&\multirow{3}{*}{$\Gamma_{\rm EM}=0.70^{+6.20}_{-0.70}$}&\multirow{5}{*}{--}
\\ \cline{5-7}
&&&&\multirow{2}{*}{$\Xi_b({3\over2}^-)\to\Xi_b^{*}\pi$}&$g^S=5.73^{+2.94}_{-3.21}$&$330^{+2000}_{-~330} $&\multirow{3}{*}{$\Gamma_{\rm Total}=330^{+2000}_{-~330}$}&
\\&&&&&$g^D=1.40^{+0.73}_{-0.49}$&$0.0^{+2.0}_{-0.0}$&
\\ \cline{5-7}
&&&&$\Xi_b({3\over2}^-)\to\Xi_b \rho\to\Xi_b \pi \pi$&$g^S=1.86^{+0.68}_{-0.52}$&$0.0^{+1.0}_{-0.0}$&
\\ \hline\hline
\multirow{3}{*}{$[\mathbf{\bar 3}_F,0,1,\rho]$}&\multirow{1}{*}{$\Lambda_b({1\over2}^-)$}&\multirow{1}{*}{$5.92^{+0.17}_{-0.19}$}
&--&--&--&--&--&--
\\ \cline{2-9}
&\multirow{2}{*}{$\Xi_b({1\over2}^-)$}&\multirow{2}{*}{$6.10^{+0.08}_{-0.08}$}
&\multirow{2}{*}{--}&\multirow{2}{*}{$\Xi_b({1\over2}^-)\to \Xi_b\pi$}
&\multirow{2}{*}{$g^S=13.90^{+5.00}_{-6.57}$}&\multirow{2}{*}{$12000^{+10000}_{-~9300}$}&\multirow{1}{*}{$\Gamma_{\rm EM}=12.0^{+10.0}_{-~9.0}$}&\multirow{2}{*}{--}
\\&&&&&&&$\Gamma_{\rm Total}=12000^{+10000}_{-~9300}$
\\ \hline\hline
\multirow{12}{*}{$[\mathbf{\bar 3}_F,1,1,\rho]$}&\multirow{3}{*}{$\Lambda_b({1\over2}^-)$}&\multirow{3}{*}{$5.92^{+0.13}_{-0.10}$}
&\multirow{6}{*}{$8\pm3$}&$\Lambda_b({1\over2}^-)\to \Sigma_b\pi\to\Lambda_b\pi\pi$&$g^S=0.84^{+1.49}_{-0.51}$&$ 2.0^{+13.0}_{-~2.0}\times 10^{-3}$&\multirow{2}{*}{$\Gamma_{\rm EM}=0.03^{+0.08}_{-0.03}$}
&\multirow{3}{*}{$\Lambda_b(5912)^0$}
\\ \cline{5-7}
&&&&$\Lambda_b({1\over2}^-)\to\Sigma_b^*\pi\to\Lambda_b\pi\pi$&$g^D=0.49^{+1.14}_{-0.24}$&$2.3^{+22.0}_{-~2.3}\times 10^{-8}$&\multirow{2}{*}{$\Gamma_{\rm Total}=0.03^{+0.09}_{-0.03}$}&
\\ \cline{5-7}
&&&&$\Lambda_b({1\over2}^-)\to \Lambda_b\rho\to\Lambda_b\pi\pi$&$g^S=1.97^{+3.20}_{-1.02}$&$1.4^{+8.4}_{-1.1}\times10^{-4}$&&
\\ \cline{2-3} \cline{5-9}
&\multirow{3}{*}{$\Lambda_b({3\over2}^-)$}&\multirow{3}{*}{$5.92^{+0.13}_{-0.10}$}&
&\multirow{2}{*}{$\Lambda_b({3\over2}^-)\to\Sigma_b^*\pi\to\Lambda_b\pi\pi$}&$g^S=0.80^{+1.54}_{-0.45}$
&$4.7^{+29.0}_{-~3.8}\times 10^{-3}$&
\\&&&&&$g^D=0.79^{+0.33}_{-0.38}$&$9.7^{+26.0}_{-~5.9}\times 10^{-6}$
&\multirow{1}{*}{$\Gamma_{\rm EM}=0.06^{+0.14}_{-0.05}$}&\multirow{2}{*}{$\Lambda_b(5920)^0$}
\\ \cline{5-7}
&&&&$\Lambda_b({3\over2}^-)\to\Sigma_b\pi$&$g^D=1.36^{+0.61}_{-0.66}$&$1.7^{+2.1}_{-1.2}\times 10^{-3}$&\multirow{1}{*}{$\Gamma_{\rm Total}=0.06^{+0.17}_{-0.05}$}&
\\ \cline{5-7}
&&&&$\Lambda_b({3\over2}^-)\to \Lambda_b\rho\to\Lambda\pi \pi$ & $g^S=0.98 ^{+0.97}_{-0.98}$&$ 4.9^{+14.0}_{-~4.9}\times 10^{-5}$&&
\\ \cline{2-9}
&\multirow{3}{*}{$\Xi_b({1\over2}^-)$}&\multirow{3}{*}{$6.09^{+0.13}_{-0.12}$}
&\multirow{7}{*}{$7\pm3$}&$\Xi_b({1\over2}^-)\to \Xi_b^{\prime}\pi$&$g^S= 0.52^{+1.02}_{-0.34} $&$4.0^{+26.0}_{-~4.0}$&\multirow{2}{*}{$\Gamma_{\rm EM}=0.05^{+0.10}_{-0.05}$}
&\multirow{3}{*}{$\Xi_b(6087)^0$}
\\ \cline{5-7}
&&&&$\Xi_b({1\over2}^-)\to\Xi_b^{*}\pi\to\Xi_b\pi\pi$&$g^D= 0.60^{+0.66}_{-0.31}$&$3.0^{+10.0}_{-~2.1}\times 10^{-9}$&\multirow{2}{*}{$\Gamma_{\rm Total}=4.0^{+26.0}_{-~4.0}$}
\\ \cline{5-7}
&&&&$\Xi_b({1\over2}^-)\to\Xi_b\rho\to\Xi_b\pi\pi$&$g^S=1.40^{+2.41}_{-0.66}$&$1.9^{+12.0}_{-~1.4}\times 10^{-4}$&
\\ \cline{2-3} \cline{5-9}
&\multirow{3}{*}{$\Xi_b({3\over2}^-)$}&\multirow{3}{*}{$6.09^{+0.12}_{-0.12}$}&
&$\Xi_b({3\over2}^-)\to\Xi_b^{\prime}\pi$&$g^D=1.02^{+0.57}_{-0.46}$&$ 2.4^{+3.4}_{-1.7}\times 10^{-4}$&\multirow{3}{*}{$\Gamma_{\rm EM}=0.09^{+0.17}_{-0.09}$}&\multirow{3}{*}{$\Xi_b(6095)^0$}
\\ \cline{5-7}
&&&&\multirow{2}{*}{$\Xi_b({3\over2}^-)\to\Xi_b^{*}\pi$}&$g^S=0.50^{+1.03}_{-0.30}$&$ 0.6^{+5.4}_{-0.6}$&\multirow{3}{*}{$\Gamma_{\rm Total}=0.70^{+5.60}_{-0.70}$}&\multirow{3}{*}{$\Xi_b(6100)^-$}
\\&&&&&$g^D= 0.59^{+0.32}_{-0.27}$&$6.0^{+8.0}_{-4.2}\times 10^{-5}$&
\\ \cline{5-7}
&&&&$\Xi_b({3\over2}^-)\to\Xi_b \rho\to\Xi_b \pi \pi$&$g^S=0.60^{+0.64}_{-0.60}$&$4.1^{+14.0}_{-~4.1}\times 10^{-5}$&
\\ \hline\hline
\multirow{12}{*}{$[\mathbf{\bar 3}_F,2,1,\rho]$}&\multirow{3}{*}{$\Lambda_b({3\over2}^-)$}&\multirow{2}{*}{$5.93^{+0.13}_{-0.13}$}
&\multirow{4}{*}{$17\pm8$}&$\Lambda_b({3\over2}^-)\to \Sigma_b\pi$&$g^D=2.75^{+1.31}_{-1.26}$&$3.6^{+4.0}_{-3.6}\times 10^{-3}$&\multirow{2}{*}{$\Gamma_{\rm EM}=0.08^{+0.56}_{-0.08}$}&\multirow{3}{*}{--}
\\ \cline{5-7}
&&&&\multirow{2}{*}{$\Lambda_b({3\over2}^-)\to\Sigma_b^*\pi\to\Lambda_b\pi\pi$}&$g^S=0.38^{+0.56}_{-0.35}$
&$0.0^{+11.0}_{-~0.0}$&\multirow{2}{*}{$\Gamma_{\rm Total}=0.0^{+11.0}_{-~0.0}$}
\\&&&&&$g^D=1.58^{+0.78}_{-0.69}$&$1.7^{+2.1}_{-1.7}\times 10^{-4}$&&
\\ \cline{2-3} \cline{5-9}
&\multirow{2}{*}{$\Lambda_b({5\over2}^-)$}&\multirow{2}{*}{$5.94^{+0.13}_{-0.13}$}&
&$\Lambda_b({5\over2}^-)\to\Sigma_b\pi$&$g^D=2.59^{+1.25}_{-2.59}$&$9.0^{+4.3}_{-9.0}\times 10^{-5}$
&\multirow{1}{*}{$\Gamma_{\rm EM}=0.01^{+0.08}_{-0.01}$}&\multirow{2}{*}{--}
\\ \cline{5-7}
&&&&$\Lambda_b({5\over2}^-)\to\Sigma_b^*\pi$&$g^D=0.25^{+0.12}_{-0.25}$&$1.5^{+1.8}_{-1.5}\times 10^{-6} $&\multirow{1}{*}{$\Gamma_{\rm Total}=0.01^{+0.08}_{-0.01}$}&
\\ \cline{2-9}
&\multirow{4}{*}{$\Xi_b({3\over2}^-)$}&\multirow{4}{*}{$6.10^{+0.15}_{-0.10}$}
&\multirow{7}{*}{$14\pm7$}&
\multirow{2}{*}{$\Xi_b({3\over2}^-)\to\Xi_b^{\star}\pi$}&$g^S=0.23^{+0.20}_{-0.20}$&$0.40^{+3.00}_{-0.40}$
&\multirow{3}{*}{$\Gamma_{\rm EM}=0.15^{+0.55}_{-0.15}$}&\multirow{4}{*}{--}
\\&&&&&$g^D=1.09^{+0.49}_{-0.45}$&$0.00^{+0.50}_{-0.00}$&
\\ \cline{5-7}
&&&&$\Xi_b({3\over2}^-)\to\Xi_b\pi$&$g^D=3.07^{+1.03}_{-0.89}$&$1.0^{+9.0}_{-1.0}$
&\multirow{1}{*}{$\Gamma_{\rm Total}=1.4^{+12.0}_{-~1.4}$}
\\ \cline{5-7}
&&&&$\Xi_b({3\over2}^-)\to\Xi_b^{\prime}\pi$&$g^D=1.89^{+0.84}_{-0.77}$&$0.00^{+0.40}_{-0.00} $&
\\ \cline{5-9}
&\multirow{3}{*}{$\Xi_b({5\over2}^-)$}&\multirow{3}{*}{$6.11^{+0.15}_{-0.10}$}&
&$\Xi_b({5\over2}^-)\to\Xi_ b\pi$&$g^D=4.35^{+1.44}_{-1.27}$&$1.0^{+7.0}_{-1.0}$&\multirow{2}{*}{$\Gamma_{\rm EM}=0.02^{+0.07}_{-0.02}$}&\multirow{3}{*}{--}
\\ \cline{5-7}
&&&&$\Xi_b({5\over2}^-)\to\Xi_b^{\prime}\pi$&$g^D=1.79^{+0.77}_{-0.70}$&$0.00^{+0.20}_{-0.00} $&\multirow{2}{*}{$\Gamma_{\rm Total}=1.0^{+7.0}_{-1.0}$}
\\ \cline{5-7}
&&&&$\Xi_b({5\over2}^-)\to\Xi_b^{*}\pi$&$g^D= 0.17^{+0.08}_{-0.08}$&$0.00^{+0.01}_{-0.00}$&
\\ \hline\hline
\end{tabular}}
\label{tab:decayb3f}
\end{center}
\end{table*}
%
\section{Discussions and summary}\label{sec:summary}

In this paper we have studied strong decay properties of the $P$-wave bottom baryons belonging to the $SU(3)$  flavor antitriplet  representation through the light-cone sum rule method within the framework of heavy quark effective theory. We have systematically calculated their $D$-wave decays into ground-state bottom baryons with light pseudoscalar mesons and $S$-wave decays into ground-state bottom baryons with light vector mesons. The extracted decay widths are presented in Table~\ref{tab:decayb3f} for the four $SU(3)$  flavor antitriplet  bottom baryon multiplets, $[\mathbf{\bar3}_F, 1, 0, \lambda]$, $[\mathbf{\bar3}_F, 0, 1, \rho]$, $[\mathbf{\bar3}_F, 1, 1, \rho]$, and $[\mathbf{\bar3}_F, 2, 1, \rho]$.

In addition, their mass spectrum has been calculated in Ref.~\cite{Mao:2015gya}, their $S$-wave decays into ground-state bottom baryons with light pseudoscalar mesons have been calculated in Ref.~\cite{Tan:2023opd}, and their radiative decays have been calculated in Ref.~\cite{Luo:2024jov}. Altogether, a rather complete investigation has been performed on the $P$-wave bottom baryons of the $SU(3)$  flavor antitriplet, from which we can understand them as a whole:
\begin{itemize}

\item The $[\mathbf{\bar 3}_F, 1, 0, \lambda]$ doublet contains four excited bottom baryons, $\Lambda_b({1/2}^-)$, $\Lambda_b({3/2}^-)$, $\Xi_b({1/2}^-)$, and $\Xi_b({3/2}^-)$. The former two $\Lambda_b$ baryons can be used to explain the $\Lambda_b(5912)^0$ and $\Lambda_b(5920)^0$, but the latter two $\Xi_b$ baryons are not easy to explain the $\Xi_b(6087)^0$ and $\Xi_b(6095)^0/\Xi_b(6100)^-$. Note that the large uncertainties in our theoretical results arise from the fact that the relevant phase spaces are highly uncertain, {\it e.g.}, the mass of $\Xi_b({1/2}^-)$ is taken from Table~\ref{tabmass} as $6.10^{+0.20}_{-0.10}$~GeV, leading to a difference of 73~MeV$\sim$ 227 MeV from the $\Xi_b^{\prime}\pi$ threshold, which can substantially impact the relevant phase space.

\item The $[\mathbf{\bar 3}_F, 0, 1, \rho]$ singlet contains two excited bottom baryons, $\Lambda_b({1/2}^-)$ and $\Xi_b({1/2}^-)$. These two baryons are not easy to be observed in experiments.

\item The $[\mathbf{\bar 3}_F, 1, 1, \rho]$ doublet contains four excited bottom baryons, $\Lambda_b({1/2}^-)$, $\Lambda_b({3/2}^-)$, $\Xi_b({1/2}^-)$, and $\Xi_b({3/2}^-)$. We can use this doublet to well explain the four excited bottom baryons $\Lambda_b(5912)^0$, $\Lambda_b(5920)^0$, $\Xi_b(6087)^0$, and $\Xi_b(6095)^0/\Xi_b(6100)^-$ as a whole.

\item The $[\mathbf{\bar 3}_F, 2, 1, \rho]$ doublet contains four excited bottom baryons, $\Lambda_b({3/2}^-)$, $\Lambda_b({5/2}^-)$, $\Xi_b({3/2}^-)$, and $\Xi_b({5/2}^-)$. We can use this doublet to further predict two $\Lambda_b$ baryons and two $\Xi_b$ baryons. We refer to Ref.~\cite{Tan:2023opd} for more discussions on this prediction. Since this is a $\rho$-mode doublet, its experimental search can verify not only our approach, but also the existence of the $\rho$-mode.

\end{itemize}
Summarizing this paper, we systematically investigate the $P$-wave bottom baryons of the $SU(3)$ flavor antitriplet to arrive at four $\Lambda_b$ and four $\Xi_b$ baryons, which have limited widths (less than $100$~MeV), and, therefore are capable of being observed in experiments. Their masses, mass splittings within the same multiplets, and strong/radiative decay properties are summarized in Table~\ref{tab:decayb3f} for future experimental searching.

%
\section*{Acknowledgments}

This project is supported by
the National Natural Science Foundation of China under Grants No.~12005172, No.~12075019, and No.~12375132,
the Jiangsu Provincial Double-Innovation Program under Grant No.~JSSCRC2021488,
and
the Fundamental Research Funds for the Central Universities.

\appendix

\begin{widetext}
\section{Sum rule equations}
\label{sec:othersumrule}
In this appendix we list the sum rule equations derived in the present study. For completeness, the sum rule equations derived in Ref.~\cite{Tan:2023opd} are also listed here.

The sum rule equation for the $\Lambda_b^+({1/2}^-)$ belonging to $[\mathbf{\bar 3}_F, 1, 0, \lambda]$ is
\begin{eqnarray*}
&& G_{\Lambda_b^+[{1\over2}^-] \rightarrow \Sigma_b^{+}\pi^-} (\omega, \omega^\prime)
= { g_{\Lambda_b^+[{1\over2}^-] \rightarrow \Sigma_b^{+}\pi^-} f_{\Lambda_b^0[{1\over2}^-]} f_{\Sigma_b^{+}} \over (\bar \Lambda_{\Lambda_b^0[{1\over2}^-]} - \omega^\prime) (\bar \Lambda_{\Sigma_b^{+}} - \omega)}
\\ \nonumber &=& \int_0^\infty dt \int_0^1 du e^{i (1-u) \omega^\prime t} e^{i u \omega t} \times 8 \Big (
 \frac{3 f_\pi m_\pi^2}{4 \pi^2 t^4 (m_u + m_d)} \phi_{3;\pi}^p(u) - \frac{i f_\pi m_\pi^2 v \cdot q}{8 \pi^2 t^3 ( m_u + m_d )} \phi_{3;\pi}^\sigma(u) + \frac{i f_\pi}{16 t v \cdot q} \langle \bar q q \rangle \psi_{4;\pi}(u)
 \\ \nonumber &&+ \frac{i f_\pi t}{256 v\cdot q} \langle g_s \bar q \sigma G q\rangle \psi_{4;\pi}(u) \Big ) \, .
\end{eqnarray*}

The sum rule equation for the $\Lambda_b^0({3/2}^-)$ belonging to $[\mathbf{\bar 3}_F, 1, 0, \lambda]$ is
\begin{eqnarray*}
&& G_{\Lambda_b^0[{3\over2}^-] \rightarrow \Sigma_b^{*+}\pi^-} (\omega, \omega^\prime)
= { g_{\Lambda_b^0[{3\over2}^-] \rightarrow \Sigma_b^{*+}\pi^-} f_{\Lambda_b^0[{3\over2}^-]} f_{\Sigma_b^{*+}} \over (\bar \Lambda_{\Lambda_b^0[{3\over2}^-]} - \omega^\prime) (\bar \Lambda_{\Sigma_b^{*+}} - \omega)}
\\ \nonumber &=& \int_0^\infty dt \int_0^1 du e^{i (1-u) \omega^\prime t} e^{i u \omega t} \times 8 \Big (
 - \frac{f_\pi m_\pi^2\phi_{3;\pi}^p(u)}{6 \pi^2 t^4 (m_u + m_d)}  + \frac{i f_\pi m_\pi^2 v \cdot q}{36 \pi^2 t^3 ( m_u + m_d )} \phi_{3;\pi}^\sigma(u) - \frac{i f_\pi}{72 t v \cdot q} \langle \bar q q \rangle \psi_{4;\pi}(u)
 \\ \nonumber &&- \frac{i f_\pi t}{1152 v\cdot q} \langle g_s \bar q \sigma G q\rangle \psi_{4;\pi}(u) \Big ) \, .
\end{eqnarray*}

The sum rule equations for the $\Xi_b^-({1/2}^-)$ belonging to $[\mathbf{\bar 3}_F, 1, 0, \lambda]$ are
\begin{eqnarray*}
&& G_{\Xi_b^-[{1\over2}^-] \rightarrow \Xi_b^{\prime0}\pi^-} (\omega, \omega^\prime)
= { g_{\Xi_b^-[{1\over2}^-] \rightarrow \Xi_b^{\prime0}\pi^-} f_{\Xi_b^-[{1\over2}^-]} f_{\Xi_b^{\prime0}} \over (\bar \Lambda_{\Xi_b^-[{1\over2}^-]} - \omega^\prime) (\bar \Lambda_{\Xi_b^{\prime0}} - \omega)}
\\ \nonumber &=& \int_0^\infty dt \int_0^1 du e^{i (1-u) \omega^\prime t} e^{i u \omega t}  \times4 \Big (
 \frac{3 f_\pi m_\pi^2}{4 \pi^2 t^4 (m_u + m_d)} \phi_{3;\pi}^p(u) - \frac{i f_\pi m_\pi^2 v \cdot q}{8 \pi^2 t^3 ( m_u + m_d )} \phi_{3;\pi}^\sigma(u) + \frac{i f_\pi}{16 t v \cdot q} \langle \bar s s \rangle \psi_{4;\pi}(u)
 \\ \nonumber &&+ \frac{i f_\pi t}{256 v\cdot q} \langle g_s \bar s \sigma G s\rangle \psi_{4;\pi}(u)
 + \frac{3 i f_\pi}{16 \pi^2 t^3 v\cdot q} m_s \psi_{4;\pi}(u) + \frac{ f_\pi m_\pi^2 m_s}{32 (m_u + m_d)} \langle \bar s s \rangle \phi_{3;\pi}^p(u) - \frac{i f_\pi m_\pi^2 t v\cdot q  m_s}{192 ( m_u + m_d)} \langle \bar s s \rangle \phi_{3;\pi}^\sigma(u) \Big ) \,
\\
\\ && G_{\Xi_b^{-}[{1\over2}^-] \rightarrow \Xi_b^{0}\rho^-} (\omega, \omega^\prime)
= { g_{\Xi_b^{-}[{1\over2}^-] \rightarrow \Xi_b^{0}\rho^-} f_{\Xi_b^{-}[{1\over2}^-]} f_{\Xi_b^{0}} \over (\bar \Lambda_{\Xi_b^{-}[{1\over2}^-]} - \omega^\prime) (\bar \Lambda_{\Xi_b^{0}} - \omega)}
\\ \nonumber &=& \int_0^\infty dt \int_0^1 du e^{i (1-u) \omega^\prime t} e^{i u \omega t} \times 4 \Big (
 \frac{i f_{\rho}^\parallel m_{\rho}}{4 \pi^2 t^4} \phi_{2;\rho}^\parallel(u)
- \frac{i f_{\rho}^\parallel m_{\rho}^3}{8 \pi^2 t^4 (v\cdot q)^2} \phi_{2;\rho}^\parallel(u)
+ \frac{i f_{\rho}^\parallel m_{\rho}^3}{4 \pi^2 t^4 (v\cdot q)^2} \phi_{3;\rho}^\perp(u)
\\ \nonumber &&- \frac{i f_{\rho}^\parallel m_{\rho}^3}{8 \pi^2 t^4 (v\cdot q)^2} \psi_{4;\rho}^\parallel(u)
+ \frac{i f_{\rho}^\parallel m_{\rho}^3}{64 \pi^2 t^2} \phi_{4;\rho}^\parallel(u)
- \frac{i f_{\rho}^\perp m_{\rho}^2}{48} \langle \bar s s \rangle \psi_{3;\rho}^\parallel(u)
- \frac{i f_{\rho}^\perp m_{\rho}^2 t^2}{768} \langle g_s \bar s \sigma G s \rangle \psi_{3;\rho}^\parallel(u)
- \frac{i f_{\rho}^\perp m_{\rho}^2}{16 \pi^2 t^2} m_s \psi_{3;\rho}^\parallel(u)
\\ \nonumber &&+ \frac{i f_{\rho}^\parallel m_{\rho}}{96} m_s \langle \bar s s \rangle \phi_{2;\rho}^\parallel(u)
- \frac{i f_{\rho}^\parallel m_{\rho}^3}{192 (v\cdot q)^2} m_s \langle \bar s s \rangle \phi_{2;\rho}^\parallel(u)
+ \frac{i f_{\rho}^\parallel m_{\rho}^3}{96 (v\cdot q)^2} m_s \langle \bar s s \rangle \phi_{3;\rho}^\perp(u)
+ \frac{i f_{\rho}^\parallel m_{\rho}^3 t^2}{1536} m_s \langle \bar s s \rangle \phi_{4;\rho}^\parallel(u)
\\ \nonumber &&
- \frac{i f_{\rho}^\parallel m_{\rho}^3}{192 (v\cdot q)^2} m_s \langle \bar s s \rangle \psi_{4;\rho}^\parallel(u)
 \Big )+ \int_0^\infty dt \int_0^1 du \int \mathcal{D}\underline{\alpha} e^{i \omega^\prime t (\alpha_2 + u\alpha_3)} e^{i \omega t(1-\alpha_2-u\alpha_3)} \times 4 \times \Big (
 - \frac{i f_{\rho}^\parallel m_{\rho}^3}{8 \pi^2 t^2} \Phi_{4;\rho}^\parallel(\underline{\alpha})
 \\ \nonumber &&
- \frac{i f_{\rho}^\parallel m_{\rho}^3}{16 \pi^2 t^2} \Psi_{4;\rho}^\parallel(\underline{\alpha})
- \frac{ i f_{\rho}^\parallel m_{\rho}^3}{8 \pi^2 t^2} \widetilde \Phi_{4;\rho}^\parallel(\underline{\alpha}) - \frac{i f_{\rho}^\parallel m_{\rho}^3}{16 \pi^2 t^2} \widetilde \Psi_{4;\rho}^\parallel(\underline{\alpha})
- \frac{i f_{\rho}^\parallel m_{\rho}^3 u}{4 \pi^2 t^2} \Phi_{4;\rho}^\parallel(\underline{\alpha}) - \frac{i f_{\rho}^\parallel m_{\rho}^3 u}{8 \pi^2 t^2} \Psi_{4;\rho}^\parallel(\underline{\alpha}) \Big ) \, .
\end{eqnarray*}

The sum rule equations for the $\Xi_b^-({3/2}^-)$ belonging to $[\mathbf{\bar 3}_F, 1, 0, \lambda]$ are
\begin{eqnarray*}
&& G_{\Xi_b^-[{3\over2}^-] \rightarrow \Xi_b^{*0}\pi^-} (\omega, \omega^\prime)
= { g_{\Xi_b^-[{3\over2}^-] \rightarrow \Xi_b^{*0}\pi^-} f_{\Xi_b^-[{3\over2}^-]} f_{\Xi_b^{*0}} \over (\bar \Lambda_{\Xi_b^-[{3\over2}^-]} - \omega^\prime) (\bar \Lambda_{\Xi_b^{*0}} - \omega)}
\\ \nonumber &=& \int_0^\infty dt \int_0^1 du e^{i (1-u) \omega^\prime t} e^{i u \omega t} \times 4 \Big (
 - \frac{f_\pi m_\pi^2 \phi_{3;\pi}^p(u)}{6 \pi^2 t^4 (m_u + m_d)}  + \frac{i f_\pi m_\pi^2 v \cdot q}{36 \pi^2 t^3 ( m_u + m_d )} \phi_{3;\pi}^\sigma(u)
 - \frac{i f_\pi}{72 t v \cdot q} \langle \bar s s \rangle \psi_{4;\pi}(u)
 \\ \nonumber &&- \frac{i f_\pi t\psi_{4;\pi}(u)}{1152 v\cdot q} \langle g_s \bar s \sigma G s\rangle
 - \frac{i f_\pi m_s}{24 \pi^2 t^3 v\cdot q} \psi_{4;\pi}(u) - \frac{ f_\pi m_\pi^2 m_s}{144 (m_u + m_d)} \langle \bar s s \rangle \phi_{3;\pi}^p(u) + \frac{i f_\pi m_\pi^2  m_s t v\cdot q}{864 ( m_u + m_d)} \langle \bar s s \rangle \phi_{3;\pi}^\sigma(u) \Big ) \, ,
\\
&& G_{\Xi_b^{-}[{3\over2}^-] \rightarrow \Xi_b^{0}\rho^-} (\omega, \omega^\prime)
= { g_{\Xi_b^{-}[{3\over2}^-] \rightarrow \Xi_b^{0}\rho^-} f_{\Xi_b^{-}[{3\over2}^-]} f_{\Xi_b^{0}} \over (\bar \Lambda_{\Xi_b^{-}[{3\over2}^-]} - \omega^\prime) (\bar \Lambda_{\Xi_b^{0}} - \omega)}
\\ \nonumber &=& \int_0^\infty dt \int_0^1 du e^{i (1-u) \omega^\prime t} e^{i u \omega t} \times 4 \Big (
 \frac{i f_{\rho}^\parallel m_{\rho}}{6 \pi^2 t^4} \phi_{2;\rho}^\parallel(u)
- \frac{i f_{\rho}^\parallel m_{\rho}^3}{12 \pi^2 t^4 (v\cdot q)^2} \phi_{2;\rho}^\parallel(u)
+ \frac{i f_{\rho}^\parallel m_{\rho}^3}{6 \pi^2 t^4 (v\cdot q)^2} \phi_{3;\rho}^\perp(u)
\\ \nonumber &&- \frac{i f_{\rho}^\parallel m_{\rho}^3}{12 \pi^2 t^4 (v\cdot q)^2} \psi_{4;\rho}^\parallel(u)
+ \frac{i f_{\rho}^\parallel m_{\rho}^3}{96 \pi^2 t^2} \phi_{4;\rho}^\parallel(u)
- \frac{i f_{\rho}^\perp m_{\rho}^2}{72} \langle \bar s s \rangle \psi_{3;\rho}^\parallel(u)
- \frac{i f_{\rho}^\perp m_{\rho}^2 t^2}{1152} \langle g_s \bar s \sigma G s \rangle \psi_{3;\rho}^\parallel(u)
- \frac{i  m_s f_{\rho}^\perp m_{\rho}^2}{24 \pi^2 t^2}  \psi_{3;\rho}^\parallel(u)
\\ \nonumber &&+ \frac{i f_{\rho}^\parallel m_{\rho} m_s }{144}\langle \bar s s \rangle \phi_{2;\rho}^\parallel(u)
- \frac{i f_{\rho}^\parallel m_{\rho}^3 m_s}{288 (v\cdot q)^2} \langle \bar s s \rangle \phi_{2;\rho}^\parallel(u)
+ \frac{i f_{\rho}^\parallel m_{\rho}^3m_s}{144 (v\cdot q)^2}  \langle \bar s s \rangle \phi_{3;\rho}^\perp(u)
+ \frac{i f_{\rho}^\parallel m_{\rho}^3 t^2 m_s }{2304}\langle \bar s s \rangle \phi_{4;\rho}^\parallel(u)
\\ \nonumber &&
- \frac{i f_{\rho}^\parallel m_{\rho}^3 m_s}{288 (v\cdot q)^2} \langle \bar s s \rangle \psi_{4;\rho}^\parallel(u)
 \Big )
+ \int_0^\infty dt \int_0^1 du \int \mathcal{D}\underline{\alpha} e^{i \omega^\prime t (\alpha_2 + u\alpha_3)} e^{i \omega t(1-\alpha_2-u\alpha_3)} \times 4 \Big (
 - \frac{i f_{\rho}^\parallel m_{\rho}^3}{12 \pi^2 t^2} \Phi_{4;\rho}^\parallel(\underline{\alpha})
 \\ \nonumber &&
 - \frac{i f_{\rho}^\parallel m_{\rho}^3}{24 \pi^2 t^2} \Psi_{4;\rho}^\parallel(\underline{\alpha})
- \frac{ i f_{\rho}^\parallel m_{\rho}^3}{12 \pi^2 t^2} \widetilde \Phi_{4;\rho}^\parallel(\underline{\alpha}) - \frac{i f_{\rho}^\parallel m_{\rho}^3}{24 \pi^2 t^2} \widetilde \Psi_{4;\rho}^\parallel(\underline{\alpha})
- \frac{i f_{\rho}^\parallel m_{\rho}^3 u}{6 \pi^2 t^2} \Phi_{4;\rho}^\parallel(\underline{\alpha}) - \frac{i f_{\rho}^\parallel m_{\rho}^3 u}{12 \pi^2 t^2} \Psi_{4;\rho}^\parallel(\underline{\alpha}) \Big ) \, .
\end{eqnarray*}

The sum rule equation for the $\Xi_b^-({1/2}^-)$ belonging to $[\mathbf{\bar 3}_F, 0, 1, \rho]$ is
\begin{eqnarray*}
&& G_{\Xi_b^-[{1\over2}^-] \rightarrow \Xi_b^{0}\pi^-} (\omega, \omega^\prime)
= { g_{\Xi_b^-[{1\over2}^-] \rightarrow \Xi_b^{0}\pi^-} f_{\Xi_b^-[{1\over2}^-]} f_{\Xi_b^{0}} \over (\bar \Lambda_{\Xi_b^-[{1\over2}^-]} - \omega^\prime) (\bar \Lambda_{\Xi_b^{0}} - \omega)}
\\ \nonumber &=& \int_0^\infty dt \int_0^1 du e^{i (1-u) \omega^\prime t} e^{i u \omega t} \times 4 \Big (
 - \frac{3 f_\pi m_\pi^2}{4 \pi^2 t^4 (m_u + m_d)} \phi_{3;\pi}^p(u) + \frac{i f_\pi m_\pi^2 v \cdot q}{8 \pi^2 t^3 ( m_u + m_d )} \phi_{3;\pi}^\sigma(u) - \frac{i f_\pi}{16 t v \cdot q} \langle \bar s s \rangle \psi_{4;\pi}(u)
 \\ \nonumber &&- \frac{i f_\pi t}{256 v\cdot q} \langle g_s \bar s \sigma G s\rangle \psi_{4;\pi}(u)
- \frac{3 i f_\pi  m_s}{16 \pi^2 t^3 v\cdot q} \psi_{4;\pi}(u) - \frac{ f_\pi m_\pi^2 m_s}{32 (m_u + m_d)} \langle \bar s s \rangle \phi_{3;\pi}^p(u) + \frac{i f_\pi m_\pi^2 t v\cdot q m_s}{192 ( m_u + m_d)} \langle \bar s s \rangle \phi_{3;\pi}^\sigma(u) \Big ) \, .
\end{eqnarray*}

The sum rule equation for the $\Lambda_b^0({1/2}^-)$ belonging to $[\mathbf{\bar 3}_F, 1, 1, \rho]$ is
\begin{eqnarray*}
&& G_{\Lambda_b^0[{1\over2}^-] \rightarrow \Sigma_b^{+}\pi^-} (\omega, \omega^\prime)
= { g_{\Lambda_b^0[{1\over2}^-] \rightarrow \Sigma_b^{+}\pi^-} f_{\Lambda_b^0[{1\over2}^-]} f_{\Sigma_b^{+}} \over (\bar \Lambda_{\Lambda_b^0[{1\over2}^-]} - \omega^\prime) (\bar \Lambda_{\Sigma_b^{+}} - \omega)}
\\ \nonumber &=& \int_0^\infty dt \int_0^1 du e^{i (1-u) \omega^\prime t} e^{i u \omega t} \times 8 \Big (
\frac{3 i f_\pi v \cdot q }{2 \pi^2 t^4} \phi_{2;\pi}(u) + \frac{3 i f_\pi v \cdot q \phi_{4;\pi}(u)}{32 \pi^2 t^2} + \frac{3 i f_\pi \psi_{4;\pi}(u)}{2 \pi^2 t^4 v \cdot q}
+ \frac{i f_\pi m_\pi^2 v \cdot q}{24 (m_u + m_d)} {\langle \bar q q \rangle} \phi^\sigma_{3;\pi}(u)
\\ \nonumber && + \frac{i f_\pi m_\pi^2 t^2 v \cdot q \phi_{3;\pi}^\sigma(u)}{384 (m_u + m_d)} \langle g_s \bar q \sigma G q\rangle  \Big )
+ \int_0^\infty dt \int_0^1 du \int \mathcal{D}\underline{\alpha} e^{i \omega^\prime t (\alpha_2 + u\alpha_3)} e^{i \omega t(1-\alpha_2-u\alpha_3)} \times 8 \times \Big (
\frac{3 i f_\pi v\cdot q}{8 \pi^2 t^2} \Phi_{4;\pi}(\underline{\alpha})
\\ \nonumber &&
- \frac{i f_\pi v\cdot q}{4 \pi^2 t^2} \Psi_{4;\pi}(\underline{\alpha}) + \frac{i f_\pi v\cdot q}{8 \pi^2 t^2} \widetilde \Phi_{4;\pi}(\underline{\alpha}) + \frac{i f_\pi v\cdot q}{4 \pi^2 t^2} \widetilde \Psi_{4;\pi}(\underline{\alpha})
+ \frac{i f_\pi u v \cdot q }{4 \pi^2 t^2} \Phi_{4;\pi}(\underline{\alpha}) + \frac{i f_\pi u v\cdot q}{2 \pi^2 t^2} \Psi_{4;\pi}(\underline{\alpha}) \Big ) \, .
\end{eqnarray*}

One of the sum rule equations for the $\Lambda_b^0({3/2}^-)$ belonging to $[\mathbf{\bar 3}_F, 1, 1, \rho]$ has been given in Eq.~(\ref{eq:g}), the other is
\begin{eqnarray*}
&& G_{\Lambda_b^0[{3\over2}^-] \rightarrow \Sigma_b^{*+}\pi^-} (\omega, \omega^\prime)
= { g_{\Lambda_b^0[{3\over2}^-] \rightarrow \Sigma_b^{*+}\pi^-} f_{\Lambda_b^0[{3\over2}^-]} f_{\Sigma_b^{*+}} \over (\bar \Lambda_{\Lambda_b^0[{3\over2}^-]} - \omega^\prime) (\bar \Lambda_{\Sigma_b^{*+}} - \omega)}
\\ \nonumber &=& \int_0^\infty dt \int_0^1 du e^{i (1-u) \omega^\prime t} e^{i u \omega t} \times 8 \Big (
 - \frac{f_\pi v \cdot q \phi_{2;\pi}(u)}{3 \pi^2 t^4}  - \frac{f_\pi v \cdot q \phi_{4;\pi}(u)}{48 \pi^2 t^2} - \frac{f_\pi \psi_{4;\pi}(u)}{3 \pi^2 t^4 v \cdot q}  - \frac{f_\pi m_\pi^2 v \cdot q}{108 (m_u + m_d)} {\langle \bar q q \rangle} \phi^\sigma_{3;\pi}(u)
\\ \nonumber && - \frac{f_\pi m_\pi^2 t^2 v \cdot q}{1728 (m_u + m_d)} \langle g_s \bar q \sigma G q\rangle \phi_{3;\pi}^\sigma(u) \Big )
+ \int_0^\infty dt \int_0^1 du \int \mathcal{D}\underline{\alpha} e^{i \omega^\prime t (\alpha_2 + u\alpha_3)} e^{i \omega t(1-\alpha_2-u\alpha_3)} \times 8  \Big (
 - \frac{f_\pi v\cdot q}{24 \pi^2 t^2} \Phi_{4;\pi}(\underline{\alpha})
  \\\ \nonumber &&+ \frac{5 f_\pi v\cdot q}{72 \pi^2 t^2} \Psi_{4;\pi}(\underline{\alpha}) - \frac{f_\pi v\cdot q}{72 \pi^2 t^2} \widetilde \Phi_{4;\pi}(\underline{\alpha}) - \frac{5 f_\pi v\cdot q}{72 \pi^2 t^2} \widetilde \Psi_{4;\pi}(\underline{\alpha})
 - \frac{f_\pi u v \cdot q }{36 \pi^2 t^2} \Phi_{4;\pi}(\underline{\alpha})
- \frac{7 f_\pi u v\cdot q}{72 \pi^2 t^2} \Psi_{4;\pi}(\underline{\alpha})
 \\ \nonumber &&
 + \frac{f_\pi u v \cdot q }{12 \pi^2 t^2} \widetilde \Phi_{4;\pi}(\underline{\alpha})
 + \frac{f_\pi u v\cdot q}{24 \pi^2 t^2} \widetilde \Psi_{4;\pi}(\underline{\alpha}) \Big ) \, .
\end{eqnarray*}

The sum rule equation for the $\Xi_b^-({1/2}^-)$ belonging to $[\mathbf{\bar 3}_F, 1, 1, \rho]$ is
\begin{eqnarray*}
&& G_{\Xi_b^-[{1\over2}^-] \rightarrow \Xi_b^{0}\rho^-} (\omega, \omega^\prime)
= { g_{\Xi_b^-[{1\over2}^-] \rightarrow \Xi_b^{0}\rho^-} f_{\Xi_b^-[{1\over2}^-]} f_{\Xi_b^{0}} \over (\bar \Lambda_{\Xi_b^-[{1\over2}^-]} - \omega^\prime) (\bar \Lambda_{\Xi_b^{0}} - \omega)}
\\ \nonumber &=& \int_0^\infty dt \int_0^1 du e^{i (1-u) \omega^\prime t} e^{i u \omega t} \times 4  \Big (
- \frac{ f_{\rho}^\perp v\cdot q}{2 \pi^2 t^4} \phi_{2;\rho}^\perp(u)
+ \frac{ f_{\rho}^\perp m_\rho^2 }{2 \pi^2 t^4 v\cdot q} \phi_{2;\rho}^\perp(u)
- \frac{ f_{\rho}^\perp m_\rho^2 }{2 \pi^2 t^4 v\cdot q} \psi_{4;\rho}^\perp(u)
- \frac{f_{\rho}^\perp m_{\rho}^2 v\cdot q}{32 \pi^2 t^2} \phi_{4;\rho}^\perp(u)
\\ \nonumber &&+ \frac{f_{\rho}^\parallel m_{\rho} v\cdot q}{48} \langle \bar s s \rangle \psi_{3;\rho}^\perp(u)
+ \frac{f_{\rho}^\parallel m_{\rho} t^2 v\cdot q}{768} \langle g_s \bar s \sigma G s \rangle \psi_{3;\rho}^\perp(u)
+ \frac{ f_{\rho}^\parallel m_{\rho} v\cdot q}{16 \pi^2 t^2} m_s \psi_{3;\rho}^\perp(u)
- \frac{ f_{\rho}^\perp v\cdot q}{48} m_s \langle \bar s s \rangle \phi_{2;\rho}^\perp(u)
\\ \nonumber &&
+ \frac{ f_{\rho}^\perp m_\rho^2}{48 v\cdot q} m_s \langle \bar s s \rangle \phi_{2;\rho}^\perp(u)
- \frac{ f_{\rho}^\perp m_\rho^2}{48 v\cdot q} m_s \langle \bar s s \rangle \psi_{4;\rho}^\perp(u)
- \frac{ f_{\rho}^\perp m_{\rho}^2 t^2 v\cdot q}{768} m_s \langle \bar s s \rangle \phi_{4;\rho}^\perp(u)
 \Big )
\\ \nonumber &+& \int_0^\infty dt \int_0^1 du \int \mathcal{D}\underline{\alpha} e^{i \omega^\prime t (\alpha_2 + u\alpha_3)} e^{i \omega t(1-\alpha_2-u\alpha_3)} \times 4 \Big (
 - \frac{ f_{\rho}^\perp m_{\rho}^2 v \cdot q}{8 \pi^2 t^2} \Psi_{4;\rho}^\perp(\underline{\alpha})
+ \frac{ f_{\rho}^\perp m_{\rho}^2 v \cdot q}{8 \pi^2 t^2} \widetilde \Psi_{4;\rho}^\perp(\underline{\alpha})
\\ \nonumber &&- \frac{ f_{\rho}^\perp m_{\rho}^2 u v \cdot q}{8 \pi^2 t^2} \Phi_{4;\rho}^{\perp1}(\underline{\alpha}) + \frac{ f_{\rho}^\perp m_{\rho}^2 u v \cdot q}{8 \pi^2 t^2} \Phi_{4;\rho}^{\perp2}(\underline{\alpha})
- \frac{ f_{\rho}^\perp m_{\rho}^2 u v \cdot q}{4 \pi^2 t^2} \widetilde \Psi_{4;\rho}^{\perp}(\underline{\alpha})
 \Big ) \, .
\end{eqnarray*}

The sum rule equations for the $\Xi_b^-({3/2}^-)$ belonging to $[\mathbf{\bar 3}_F, 1, 1, \rho]$ are
\begin{eqnarray*}
&& G_{\Xi_b^-[{3\over2}^-] \rightarrow \Xi_b^{*0}\pi^-} (\omega, \omega^\prime)
= { g_{\Xi_b^-[{3\over2}^-] \rightarrow \Xi_b^{*0}\pi^-} f_{\Xi_b^-[{3\over2}^-]} f_{\Xi_b^{*0}} \over (\bar \Lambda_{\Xi_b^-[{3\over2}^-]} - \omega^\prime) (\bar \Lambda_{\Xi_b^{*0}} - \omega)}
\\ \nonumber &=& \int_0^\infty dt \int_0^1 du e^{i (1-u) \omega^\prime t} e^{i u \omega t} \times 4 \Big (- \frac{f_\pi v \cdot q \phi_{2;\pi}(u)}{3 \pi^2 t^4}  - \frac{f_\pi v \cdot q \phi_{4;\pi}(u)}{48 \pi^2 t^2}  - \frac{f_\pi \psi_{4;\pi}(u)}{3 \pi^2 t^4 v \cdot q}
- \frac{f_\pi m_\pi^2 v \cdot q {\langle \bar s s \rangle} }{108 (m_u + m_d)} \phi^\sigma_{3;\pi}(u)
\\ \nonumber &&
- \frac{f_\pi m_\pi^2 t^2 v \cdot q}{1728 (m_u + m_d)} \langle g_s \bar s \sigma G s\rangle \phi_{3;\pi}^\sigma(u)
- \frac{f_\pi m_\pi^2 v \cdot q \phi_{3;\pi}^\sigma(u)}{36 \pi^2 t^2 (m_u + m_d)} m_s  - \frac{f_\pi v \cdot q \phi_{2;\pi}(u)}{72} m_s {\langle \bar s s \rangle}
- \frac{f_\pi t^2 v\cdot q}{1152} m_s {\langle \bar s s \rangle} \phi_{4;\pi}(u)
\\ \nonumber &&- \frac{f_\pi }{72 v\cdot q} m_s {\langle \bar s s \rangle} \psi_{4;\pi}(u) \Big )
+ \int_0^\infty dt \int_0^1 du \int \mathcal{D}\underline{\alpha} e^{i \omega^\prime t (\alpha_2 + u\alpha_3)} e^{i \omega t(1-\alpha_2-u\alpha_3)} \times 4 \Big (
 - \frac{f_\pi v\cdot q}{24 \pi^2 t^2} \Phi_{4;\pi}(\underline{\alpha})
 \\ \nonumber &&
 + \frac{5 f_\pi v\cdot q}{72 \pi^2 t^2} \Psi_{4;\pi}(\underline{\alpha}) - \frac{f_\pi v\cdot q}{72 \pi^2 t^2} \widetilde \Phi_{4;\pi}(\underline{\alpha}) - \frac{5 f_\pi v\cdot q}{72 \pi^2 t^2} \widetilde \Psi_{4;\pi}(\underline{\alpha})
 - \frac{f_\pi u v \cdot q }{36 \pi^2 t^2} \Phi_{4;\pi}(\underline{\alpha}) - \frac{7 f_\pi u v\cdot q}{72 \pi^2 t^2} \Psi_{4;\pi}(\underline{\alpha}) + \frac{f_\pi u v \cdot q }{12 \pi^2 t^2} \widetilde \Phi_{4;\pi}(\underline{\alpha})
 \\ \nonumber &&+ \frac{f_\pi u v\cdot q}{24 \pi^2 t^2} \widetilde \Psi_{4;\pi}(\underline{\alpha}) \Big ) \, ,
\\
&& G_{\Xi_b^{-}[{3\over2}^-] \rightarrow \Xi_b^{0}\rho^-} (\omega, \omega^\prime)
= { g_{\Xi_b^{-}[{3\over2}^-] \rightarrow \Xi_b^{0}\rho^-} f_{\Xi_b^{-}[{3\over2}^-]} f_{\Xi_b^{0}} \over (\bar \Lambda_{\Xi_b^{-}[{3\over2}^-]} - \omega^\prime) (\bar \Lambda_{\Xi_b^{0}} - \omega)}
\\ \nonumber &=& \int_0^\infty dt \int_0^1 du e^{i (1-u) \omega^\prime t} e^{i u \omega t} \times 4 \Big (
 \frac{ f_{\rho}^\perp v\cdot q}{3 \pi^2 t^4} \phi_{2;\rho}^\perp(u)
- \frac{ f_{\rho}^\perp m_\rho^2 }{3 \pi^2 t^4 v\cdot q} \phi_{2;\rho}^\perp(u)
+ \frac{ f_{\rho}^\perp m_\rho^2 }{3 \pi^2 t^4 v\cdot q} \psi_{4;\rho}^\perp(u)
\\ \nonumber &&
+\frac{f_{\rho}^\perp m_{\rho}^2 v\cdot q}{48 \pi^2 t^2} \phi_{4;\rho}^\perp(u)
-\frac{f_{\rho}^\parallel m_{\rho} v\cdot q}{72} \langle \bar s s \rangle \psi_{3;\rho}^\perp(u)
-\frac{f_{\rho}^\parallel m_{\rho} t^2 v\cdot q}{1152} \langle g_s \bar s \sigma G s \rangle \psi_{3;\rho}^\perp(u)
- \frac{ f_{\rho}^\parallel m_{\rho} v\cdot q}{24 \pi^2 t^2} m_s \psi_{3;\rho}^\perp(u)
\\ \nonumber &&+ \frac{ f_{\rho}^\perp v\cdot q}{72} m_s \langle \bar s s \rangle \phi_{2;\rho}^\perp(u)
- \frac{ f_{\rho}^\perp m_\rho^2}{72 v\cdot q} m_s \langle \bar s s \rangle \phi_{2;\rho}^\perp(u)
+ \frac{ f_{\rho}^\perp m_\rho^2}{72 v\cdot q} m_s \langle \bar s s \rangle \psi_{4;\rho}^\perp(u)
+ \frac{ f_{\rho}^\perp m_{\rho}^2 t^2 v\cdot q}{1152} m_s \langle \bar s s \rangle \phi_{4;\rho}^\perp(u)
 \Big )
\\ \nonumber &&+ \int_0^\infty dt \int_0^1 du \int \mathcal{D}\underline{\alpha} e^{i \omega^\prime t (\alpha_2 + u\alpha_3)} e^{i \omega t(1-\alpha_2-u\alpha_3)} \times 4 \Big (
 \frac{ f_{\rho}^\perp m_{\rho}^2 v \cdot q}{12 \pi^2 t^2} \Psi_{4;\rho}^\perp(\underline{\alpha}) - \frac{ f_{\rho}^\perp m_{\rho}^2 v \cdot q}{12 \pi^2 t^2} \widetilde \Psi_{4;\rho}^\perp(\underline{\alpha})
\\ \nonumber &&
+ \frac{ f_{\rho}^\perp m_{\rho}^2 u v \cdot q}{12 \pi^2 t^2} \Phi_{4;\rho}^{\perp1}(\underline{\alpha}) - \frac{ f_{\rho}^\perp m_{\rho}^2 u v \cdot q}{12 \pi^2 t^2} \Phi_{4;\rho}^{\perp2}(\underline{\alpha})
+ \frac{ f_{\rho}^\perp m_{\rho}^2 u v \cdot q}{6 \pi^2 t^2} \widetilde \Psi_{4;\rho}^{\perp}(\underline{\alpha})
 \Big ) \, .
\end{eqnarray*}

The sum rule equation for the $\Lambda_b^0({3/2}^-)$ belonging to $[\mathbf{\bar 3}_F, 2, 1, \rho]$ is
\begin{eqnarray*}
&& G_{\Lambda_b^0[{3\over2}^-] \rightarrow \Sigma_b^{*+}\pi^-} (\omega, \omega^\prime)
= { g_{\Lambda_b^0[{3\over2}^-] \rightarrow \Sigma_b^{*+}\pi^-} f_{\Lambda_b^0[{3\over2}^-]} f_{\Sigma_b^{*+}} \over (\bar \Lambda_{\Lambda_b^0[{3\over2}^-]} - \omega^\prime) (\bar \Lambda_{\Sigma_b^{*+}} - \omega)}
\\ \nonumber &=& \int_0^\infty dt \int_0^1 du \int \mathcal{D}\underline{\alpha} e^{i \omega^\prime t (\alpha_2 + u\alpha_3)} e^{i \omega t(1-\alpha_2-u\alpha_3)} \times 8  \Big (
\frac{f_\pi v\cdot q}{24 \pi^2 t^2} \Phi_{4;\pi}(\underline{\alpha})
 + \frac{f_\pi v\cdot q}{24 \pi^2 t^2} \Psi_{4;\pi}(\underline{\alpha})
 - \frac{ f_\pi v\cdot q}{24 \pi^2 t^2} \widetilde \Phi_{4;\pi}(\underline{\alpha})
 \\ \nonumber &&
 - \frac{f_\pi v\cdot q}{24 \pi^2 t^2} \widetilde \Psi_{4;\pi}(\underline{\alpha})
- \frac{f_\pi u v \cdot q }{12 \pi^2 t^2} \Phi_{4;\pi}(\underline{\alpha}) - \frac{f_\pi u v\cdot q}{24 \pi^2 t^2} \Psi_{4;\pi}(\underline{\alpha}) + \frac{f_\pi u v \cdot q }{12 \pi^2 t^2} \widetilde \Phi_{4;\pi}(\underline{\alpha}) + \frac{f_\pi u v\cdot q}{24 \pi^2 t^2} \widetilde \Psi_{4;\pi}(\underline{\alpha}) \Big ) \, .
\end{eqnarray*}

The sum rule equation for the $\Lambda_b^-({5/2}^-)$ belonging to $[\mathbf{\bar 3}_F, 2, 1, \rho]$ is
\begin{eqnarray*}
&& G_{\Lambda_b^0[{5\over2}^-] \rightarrow \Sigma_b^{*+}\rho^-} (\omega, \omega^\prime)
= { g_{\Lambda_b^0[{5\over2}^-] \rightarrow \Sigma_b^{*+}\rho^-} f_{\Lambda_b^+[{5\over2}^-]} f_{\Sigma_b^{*+}} \over (\bar \Lambda_{\Lambda_b^0[{5\over2}^-]} - \omega^\prime) (\bar \Lambda_{\Sigma_b^{*+}} - \omega)}
\\ \nonumber &=& \int_0^\infty dt \int_0^1 du e^{i (1-u) \omega^\prime t} e^{i u \omega t} \times 8 \Big (
 -\frac{3 i f_\rho^\parallel m_\rho}{10 \pi^2 t^4} \phi_{2;\rho}^\parallel(u)
+ \frac{3 i f_\rho^\parallel m_\rho^3}{20 \pi^2 t^4 (v\cdot q)^2} \phi_{2;\rho}^\parallel(u)
+ \frac{3 i f_\rho^\parallel m_\rho }{10 \pi^2 t^4} \phi_{3;\rho}^\perp(u)
\\ \nonumber &&
- \frac{3 i f_\rho^\parallel m_\rho^3}{10 \pi^2 t^4 (v\cdot q)^2} \phi_{3;\rho}^\perp(u)
- \frac{3 f_\rho^\parallel m_\rho v\cdot q}{40 \pi^2 t^3} \psi_{3;\rho}^\perp(u)
- \frac{3 i f_\rho^\parallel m_\rho^3}{160 \pi^2 t^2} \phi_{4;\rho}^\parallel(u)
+ \frac{3 i f_\rho^\parallel m_\rho^3}{20 \pi^2 t^4 (v\cdot q)^2} \psi_{4;\rho}^\parallel(u)
\\ \nonumber &&
+ \frac{f_\rho^\perp m_\rho^2}{40 t v\cdot q} \langle \bar q q \rangle \phi_{2;\rho}^\perp(u)
+ \frac{i f_\rho^\perp m_\rho^2}{40} \langle \bar q q \rangle \psi_{3;\rho}^\parallel(u)
- \frac{f_\rho^\perp m_\rho^2}{40 t v\cdot q} \langle \bar q q \rangle \psi_{4;\rho}^\perp(u)
+ \frac{f_\rho^\perp m_\rho^2 t}{640 v\cdot q} \langle g_s \bar q \sigma G q \rangle \phi_{2;\rho}^\perp(u)
\\ \nonumber &&
+ \frac{ i f_\rho^\perp m_\rho^2 t^2}{640} \langle g_s \bar q \sigma G q \rangle \psi_{3;\rho}^\parallel(u)
- \frac{f_\rho^\perp m_\rho^2 t}{640 v\cdot q} \langle g_s \bar q \sigma G q \rangle \psi_{4;\rho}^\perp(u)
\Big )
\\ \nonumber &&+ \int_0^\infty dt \int_0^1 du \int \mathcal{D}\underline{\alpha} e^{i \omega^\prime t (\alpha_2 + u\alpha_3)} e^{i \omega t(1-\alpha_2-u\alpha_3)} \times 8  \Big (
 \frac{3 i f_\rho^\parallel m_\rho^3}{40 \pi^2 t^2} \Psi_{4;\rho}^\parallel(\underline{\alpha})
- \frac{3 i f_\rho^\parallel m_\rho^3}{40 \pi^2 t^2} \widetilde \Psi_{4;\rho}^\parallel(\underline{\alpha})
+ \frac{3 i f_\rho^\parallel m_\rho^3 u}{20 \pi^2 t^2} \Psi_{4;\rho}^\parallel(\underline{\alpha}) \Big ) \, .
\end{eqnarray*}

The sum rule equation for the $\Xi_b^-({3/2}^-)$ belonging to $[\mathbf{\bar 3}_F, 2, 1, \rho]$ is
\begin{eqnarray*}
&& G_{\Xi_b^-[{3\over2}^-] \rightarrow \Xi_b^{*0}\pi^-} (\omega, \omega^\prime)
= { g_{\Xi_b^-[{3\over2}^-] \rightarrow \Xi_b^{*0}\pi^-} f_{\Xi_b^-[{3\over2}^-]} f_{\Xi_b^{*0}} \over (\bar \Lambda_{\Xi_b^-[{3\over2}^-]} - \omega^\prime) (\bar \Lambda_{\Xi_b^{*0}} - \omega)}
\\ \nonumber &=& \int_0^\infty dt \int_0^1 du \int \mathcal{D}\underline{\alpha} e^{i \omega^\prime t (\alpha_2 + u\alpha_3)} e^{i \omega t(1-\alpha_2-u\alpha_3)} \times 4  \Big (
 \frac{f_\pi v\cdot q}{24 \pi^2 t^2} \Phi_{4;\pi}(\underline{\alpha})
 + \frac{f_\pi v\cdot q}{24 \pi^2 t^2} \Psi_{4;\pi}(\underline{\alpha})
 - \frac{ f_\pi v\cdot q}{24 \pi^2 t^2} \widetilde \Phi_{4;\pi}(\underline{\alpha})
 \\ \nonumber &&
 - \frac{f_\pi v\cdot q}{24 \pi^2 t^2} \widetilde \Psi_{4;\pi}(\underline{\alpha})
- \frac{f_\pi u v \cdot q }{12 \pi^2 t^2} \Phi_{4;\pi}(\underline{\alpha}) - \frac{f_\pi u v\cdot q}{24 \pi^2 t^2} \Psi_{4;\pi}(\underline{\alpha}) + \frac{f_\pi u v \cdot q }{12 \pi^2 t^2} \widetilde \Phi_{4;\pi}(\underline{\alpha}) + \frac{f_\pi u v\cdot q}{24 \pi^2 t^2} \widetilde \Psi_{4;\pi}(\underline{\alpha}) \Big ) \, .
\end{eqnarray*}

The sum rule equation for the $\Xi_b^-({5/2}^-)$ belonging to $[\mathbf{\bar 3}_F, 2, 1, \rho]$ is
\begin{eqnarray*}
&& G_{\Xi_b^-[{5\over2}^-] \rightarrow \Xi_c^{*+}{\rho}^-} (\omega, \omega^\prime)
= { g_{\Xi_b^-[{5\over2}^-] \rightarrow \Xi_c^{*+}{\rho}^-} f_{\Xi_c^0[{5\over2}^-]} f_{\Xi_b^{*0}} \over (\bar \Lambda_{\Xi_b^-[{5\over2}^-]} - \omega^\prime) (\bar \Lambda_{\Xi_b^{*0}} - \omega)}
\\ \nonumber &=& \int_0^\infty dt \int_0^1 du e^{i (1-u) \omega^\prime t} e^{i u \omega t} \times 8 \Big (
-\frac{3 i f_\rho^\parallel m_\rho}{10 \pi^2 t^4} \phi_{2;\rho}^\parallel(u)
+ \frac{3 i f_\rho^\parallel m_\rho^3}{20 \pi^2 t^4 (v\cdot q)^2} \phi_{2;\rho}^\parallel(u)
+ \frac{3 i f_\rho^\parallel m_\rho }{10 \pi^2 t^4} \phi_{3;\rho}^\perp(u)
\\ \nonumber &&
- \frac{3 i f_\rho^\parallel m_\rho^3}{10 \pi^2 t^4 (v\cdot q)^2} \phi_{3;\rho}^\perp(u)
- \frac{3 f_\rho^\parallel m_\rho v\cdot q}{40 \pi^2 t^3} \psi_{3;\rho}^\perp(u)
- \frac{3 i f_\rho^\parallel m_\rho^3}{160 \pi^2 t^2} \phi_{4;\rho}^\parallel(u)
+ \frac{3 i f_\rho^\parallel m_\rho^3}{20 \pi^2 t^4 (v\cdot q)^2} \psi_{4;\rho}^\parallel(u)
+ \frac{f_\rho^\perp m_\rho^2}{40 t v\cdot q} \langle \bar s s \rangle \phi_{2;\rho}^\perp(u)
\\ \nonumber &&
+ \frac{i f_\rho^\perp m_\rho^2}{40} \langle \bar s s \rangle \psi_{3;\rho}^\parallel(u)
- \frac{f_\rho^\perp m_\rho^2}{40 t v\cdot q} \langle \bar s s \rangle \psi_{4;\rho}^\perp(u)
+ \frac{f_\rho^\perp m_\rho^2 t}{640 v\cdot q} \langle g_s \bar s \sigma G s \rangle \phi_{2;\rho}^\perp(u)
+ \frac{ i f_\rho^\perp m_\rho^2 t^2}{640} \langle g_s \bar s \sigma G s \rangle \psi_{3;\rho}^\parallel(u)
\\ \nonumber &&
- \frac{f_\rho^\perp m_\rho^2 t}{640 v\cdot q} \langle g_s \bar s \sigma G s \rangle \psi_{4;\rho}^\perp(u)
+ \frac{3 f_{\rho}^\perp m_{\rho}^2}{40 \pi^2 t^3 v\cdot q} m_s \phi_{2;\rho}^\perp(u)
+\frac{3 i f_{\rho}^\perp m_{\rho}^2}{40 \pi^2 t^2} m_s \psi_{3;\rho}^\parallel(u)
-\frac{3 f_{\rho}^\perp m_{\rho}^2}{40 \pi^2 t^3 v\cdot q} m_s \psi_{4;\rho}^\perp(u)
\\ \nonumber &&
-\frac{ i f_{\rho}^\parallel m_{\rho}}{80} m_s \langle \bar s s \rangle \phi_{2;\rho}^\parallel(u)
+\frac{i f_{\rho}^\parallel m_{\rho}^3}{160 (v\cdot q)^2} m_s \langle \bar s s \rangle \phi_{2;\rho}^\parallel(u)
+\frac{ i f_{\rho}^\parallel m_{\rho}}{80} m_s \langle \bar s s \rangle \phi_{3;\rho}^\perp(u)
-\frac{i f_{\rho}^\parallel m_{\rho}^3}{80 (v\cdot q)^2} m_s \langle \bar s s \rangle \phi_{3;\rho}^\perp(u)
\\ \nonumber &&
-\frac{f_{\rho}^\parallel m_{\rho} t v\cdot q}{320} m_s \langle \bar s s \rangle \psi_{3;\rho}^\perp(u)
-\frac{i f_{\rho}^\parallel m_{\rho}^3 t^2 }{1280} m_s \langle \bar s s \rangle \phi_{4;\rho}^\parallel(u)
+\frac{i f_{\rho}^\parallel m_{\rho}^3}{160 (v\cdot q)^2} m_s \langle \bar s s \rangle \psi_{4;\rho}^\parallel(u)
\Big )
\\ \nonumber &&+ \int_0^\infty dt \int_0^1 du \int \mathcal{D}\underline{\alpha} e^{i \omega^\prime t (\alpha_2 + u\alpha_3)} e^{i \omega t(1-\alpha_2-u\alpha_3)} \times 8 \Big (
\frac{3 i f_\rho^\parallel m_\rho^3}{40 \pi^2 t^2} \Psi_{4;\rho}^\parallel(\underline{\alpha})
- \frac{3 i f_\rho^\parallel m_\rho^3}{40 \pi^2 t^2} \widetilde \Psi_{4;\rho}^\parallel(\underline{\alpha})+ \frac{3 i f_\rho^\parallel m_\rho^3 u}{20 \pi^2 t^2} \Psi_{4};\rho^\parallel(\underline{\alpha}) \Big ) \, .
\end{eqnarray*}
\end{widetext}

%


\end{document}